\documentclass[11pt,a4paper]{article}
\pdfoutput=1
\usepackage{jcappub,graphicx,bm}
\usepackage[dvipsnames,table]{xcolor}

\usepackage{lmodern}
\usepackage[T1]{fontenc}
\usepackage{microtype}
\usepackage[all]{nowidow}
\usepackage{etoolbox}
\AtBeginEnvironment{subequations}{\ifhmode\unskip\fi}
\makeatletter
\gdef\@fpheader{}
\makeatother

\newcommand{\tablestyle}{\displaystyle\rule[-20pt]{0pt}{42pt}}
\newcommand{\ccell}{\cellcolor{blue!8}}
\newcommand{\ceq}[1]{%
	\setlength\abovedisplayskip{0.7\jot}
	\setlength\abovedisplayshortskip{0.7\jot}
	\colorbox{blue!8}{$\displaystyle\rule[-16pt]{0pt}{38pt}\quad #1 \quad$}%
	\setlength\belowdisplayskip{0.7\jot}
	\setlength\belowdisplayshortskip{0.7\jot}
}

\makeatletter
\newcommand{\coloneq}{%
	\mathrel{\rlap{%
	 \raisebox{0.3ex}{$\m@th\cdot$}}%
	 \raisebox{-0.3ex}{$\m@th\cdot$}}%
	 =}
\makeatother
\newcommand{\Mpl}{M_\text{Pl}} 
\newcommand{\M}{\mathcal{M}} 
\renewcommand{\vec}[1]{\bm{#1}} 
\newcommand{\GN}{G_\text{N}} 
\newcommand{\avg}[1]{{\langle #1 \rangle}} 
\newcommand{\con}{\bar{c}} 
\newcommand{\dis}{\bar{d}} 
\newcommand{\sat}{{\displaystyle\ast}} 

\newcommand{\epsPM}{\epsilon_\text{PM}} 
\newcommand{\epsLadder}[1][]{\epsilon_{\text{L}{#1}}} 
\newcommand{\epsSpin}{\epsilon_\text{S}} 

\let\d\relax
\newcommand{\dx}{\text{d}}
\newcommand{\d}[2]{\frac{\dx #1}{\dx #2}}
\newcommand{\dd}[2]{\frac{\dx^2 #1}{\dx #2^2}}
\newcommand{\pd}[2]{\frac{\partial #1}{\partial #2}}

\newcommand{\scount}[1]{{\small\textsc{[\MakeLowercase{#1}]}}}

\newcommand{\vph}[1]{{\vphantom{#1}}}

\title{Spin precession as a new window into disformal scalar fields}

\author[a]{Philippe Brax,}
\author[b]{Anne-Christine Davis,}
\author[b,c]{Scott Melville}
\author[a,b]{and Leong~Khim~Wong}

\affiliation[a]{Institut de Physique Th\'eorique, Universit\'e  Paris-Saclay, CEA, CNRS,\protect\\F-91191 Gif-sur-Yvette Cedex, France}
\affiliation[b]{DAMTP, Centre for Mathematical Sciences, University of Cambridge,\protect\\Wilberforce Road, Cambridge CB3 0WA, U.K.}
\affiliation[c]{Emmanuel College, University of Cambridge,\protect\\St Andrew's Street, Cambridge CB2 3AP, U.K.}

\emailAdd{philippe.brax@ipht.fr}
\emailAdd{a.c.davis@damtp.cam.ac.uk}
\emailAdd{scott.melville@damtp.cam.ac.uk}
\emailAdd{leong-khim.wong@ipht.fr}

\abstract{%
We launch a first investigation into how a light scalar field coupled both conformally and disformally to matter influences the evolution of spinning point-like bodies. Working directly at the level of the equations of motion, we derive novel spin-orbit and spin-spin effects accurate to leading order in a nonrelativistic and weak-field expansion. Crucially, unlike the spin-independent effects induced by the disformal coupling, which have been shown to vanish in circular binaries due to rotational symmetry, the spin-dependent effects we study here persist even in the limit of zero eccentricity, and so provide a new and qualitatively distinct way of probing these kinds of interactions. To illustrate their potential, we confront our predictions with 
spin-precession measurements from the Gravity Probe~B experiment and 
find that the resulting constraint improves upon existing bounds from 
perihelion precession by over 5~orders of magnitude. Our results therefore establish spin effects as a promising window into the disformally coupled dark~sector.
}

\begin{document}
\maketitle
\raggedbottom

\section{Introduction}
\label{sec:intro}

The inclusion of a light scalar field is a simple, yet phenomenologically interesting way to construct gravitational field theories beyond general relativity. Despite having only one new degree of freedom, these models are sufficiently flexible that they enjoy a variety of applications. In cosmology, for instance, a huge ``bestiary'' of these scalar-tensor theories have been designed to serve as dark matter or dark energy candidates; while, on smaller scales, light scalar fields have also been deployed to model deviations from Einstein's theory in the strong-field regime. (See, e.g., refs.~\cite{Clifton:2011jh, Bamba:2012cp, Joyce:2014kja, Hui:2016ltb, Urena-Lopez:2019kud, Berti:2015itd} for reviews on this extensive literature.)

For theories with a real scalar field $\phi(x)$, the scalar can generically couple to matter in two distinct ways: via a conformal interaction of the form $\sim\!\phi \, T^\mu{}_\mu$, and via a disformal interaction $\sim\!\partial_\mu\phi\,\partial_\nu\phi \, T^{\mu\nu}$ (with $T^{\mu\nu}$ denoting the energy-momentum tensor of matter). As is well known, the conformal interaction prompts the mediation of an additional long-range force (the ``fifth'' force) between matter particles, whose strength and range are now tightly constrained by local tests of gravity in both the laboratory and the Solar System~\cite{Bertotti:2003rm, Adelberger:2009zz, Burrage:2017qrf, Hofmann2018, Berge:2017ovy}. The strength of the disformal interaction, in contrast, is less so\,---\,its confrontation with a variety of experiments notwithstanding \cite{Koivisto:2008ak, Zumalacarregui:2010wj, Koivisto:2012za, vandeBruck:2013yxa, Neveu:2014vua, Sakstein:2014isa, Sakstein:2014aca, Ip:2015qsa, Sakstein:2015jca, vandeBruck:2015ida, vandeBruck:2016cnh, Kaloper:2003yf, Brax:2014vva, Brax:2015hma, Brax:2012ie, vandeBruck:2012vq, Brax:2013nsa}.

In the context of binary systems, much of the difficulty with constraining this disformal coupling can be attributed to its derivative nature, which leads to its effects being suppressed by powers of the orbital velocity. Moreover, recent work~\cite{Brax:2018bow, Brax:2019tcy, Kuntz:2019zef, Melville:2019wyy} has shown that in systems comprised of nonspinning objects, the effects that stem from this disformal interaction are also suppressed by the orbital eccentricity and, in fact, vanish for perfectly circular orbits due to rotational symmetry. This eccentricity suppression makes probing disformal effects with local Solar System measurements difficult, since all of the planets around our Sun have orbits that are very nearly circular. Similarly, because most compact binaries are expected to circularise by the time they enter the LIGO-Virgo band, the constraining power of (present) gravitational-wave observations is also hindered in this regard.\looseness=-1

That being said, a key ingredient absent from these earlier studies, but present in nature, is that of spin. We shall here address this gap in the literature by deriving, for the first time, the equations of motion for a generic system of $N$ point-like (but spinning) bodies coupled both conformally and disformally to a light scalar field. While we focus here on just the leading terms in a nonrelativistic and weak-field expansion, our results already show that the inclusion of spin effects opens up a promising new window into the disformally coupled dark sector.  Indeed, because the presence of spin can break rotational symmetry even when the orbit is circular, we find that spin-dependent disformal effects are not eccentricity suppressed and so provide us with a new and qualitatively distinct way of probing these scalar interactions.
\looseness=-1

\subsection{Summary}
\label{sec:intro_summary}

Our main contributions concern two key quantities. The trajectory~$\vec x_A(t)$ of the $A$th body is determined by its acceleration vector (or, said differently, by the force per unit mass acting on it),
\begin{equation}
	\dd{\vec x_A}{t} = \vec a_A,
\end{equation}
while the rate at which its spin~$\vec S_A(t)$ precesses is set by the angular velocity vector~$\vec\Omega_A(t)$ according to the equation
\begin{equation}
	\d{\vec S_A}{t} = \vec\Omega_A \times \vec S_A.
\label{eqn:intro_eom_spin}
\end{equation}

A scalar field introduces the corrections
${ \Delta\vec a_A^\vph{(} = \vec a_A^\vph{(} - \vec a_A^\text{(GR)} }$
and
${ \Delta\vec\Omega_A^\vph{(} = \vec\Omega_A^\vph{(} - \vec\Omega_A^\text{(GR)} }$, which cause these quantities to deviate from their general relativistic values (reviewed in section~\ref{sec:gr}). These corrections can each be organised as expansions in three parameters. First, we expand in powers of the disformal coupling constant~$\dis/\M^4$, such that
\begin{subequations}
\label{eqn:intro_expansion_ladder}
\begin{align}
	\Delta\vec a_A
	&=
	\frac{\con^2}{\Mpl^2}
	\bigg(
		\Delta\vec a_A^{(0)}
		+
		\frac{\dis}{\M^4}\Delta\vec a_A^{(1)}
		+
		\cdots
	\bigg),
	\\
	\Delta\vec\Omega_A
	&=
	\frac{\con^2}{\Mpl^2}
	\bigg(
		\Delta\vec\Omega_A^{(0)}
		+
		\frac{\dis}{\M^4}\Delta\vec\Omega_A^{(1)}
		+
		\cdots
	\bigg).
\end{align}
\end{subequations}
Also appearing on the rhs is the conformal coupling constant~$\con$. On each rung $n$ in this so-called \emph{ladder expansion}, we further expand in powers of the~spins,
\begin{subequations}
\label{eqn:intro_expansion_spin}
\begin{align}
	\Delta\vec a_A^{(n)}
	&=
	\Delta\vec a_A^{(n)}\scount{o}
	+
	\Delta\vec a_A^{(n)}\scount{so}
	+
	\Delta\vec a_A^{(n)}\scount{ss}
	+
	\cdots,
	\\
	\Delta\vec\Omega_A^{(n)}
	&=
	\Delta\vec\Omega_A^{(n)}\scount{so}
	+
	\Delta\vec\Omega_A^{(n)}\scount{ss}
	+ \cdots,
\end{align}
\end{subequations}
where (adapting the notation in ref.~\cite{Will:2018}) the orbital part~(O) is independent of spin, the spin-orbit part~(SO) leads to terms in the equations of motion that are linear in the spins, and the spin-spin part~(SS) produces terms that are bilinear in the spins.
Finally, each of the quantities on the rhs of \eqref{eqn:intro_expansion_spin} further admits a nonrelativistic, post-Newtonian~(PN) expansion in powers of the velocity.

We shall work here to leading order in the velocity and in the disformal coupling, and up to bilinear order in the spins. Already at this order, the scalar-induced acceleration $\Delta\vec a_A$ has six different contributions, while the scalar-induced precession rate $\Delta\vec\Omega_A$ has four (see~table~\ref{table:summary}).

\bgroup
\setlength{\tabcolsep}{3pt}
\begin{table}
\small\centering
\begin{tabular}{|ccc|}
\hline
& Conformal $(n=0)$ $\rule[-7pt]{0pt}{20pt}$
& Disformal $(n=1)$ $\rule[-7pt]{0pt}{20pt}$ \\
\hline
$\Delta\vec a^{(n)}\scount{o}$
&
$\tablestyle
	-\frac{m}{16\pi r^2} \vec n
$
&
$\tablestyle
	-\frac{m^2}{64\pi^2 r^5}
	\big[
		\vec a \cdot \vec r
		-
		3 (\vec n \cdot \vec v)^2
		+
		\vec v^2
	\big]
	\vec n
$
\\
$ \Delta \vec a^{(n)}\scount{so} $
&
$\tablestyle
	\frac{ [\vec v - 3(\vec n \cdot \vec v)\vec n]\times\vec S_+ }
	{16\pi r^3}
$
&
\ccell
$\tablestyle
	\frac{m}{64\pi^2 r^6}
	\bigg(
		\frac{3(\vec\ell \cdot \vec S_+) \vec n}{r}
		-
		[\vec v - 3(\vec n \cdot \vec v)]\times \vec S_+
	\bigg)
$\\
$\Delta\vec a^{(n)}\scount{ss}^i$
&
$0$
&
\ccell
$\tablestyle
	\frac{9 S_{(1}^{ij} S^{k\ell \vph{j}}_{2)} }{64\pi^2 r^7}
	\big[
		5(n^{\avg{jkq}\avg{\ell p}}
		+ n^{\avg{jk}\avg{\ell pq}}) v^{pq}
		-
		r n^{\avg{jk}\avg{\ell p}} a^p
	\big]
$
\\
\hline
$ \Delta\vec\Omega_1^{(n)}\scount{so} $
&
$\tablestyle
	- \frac{\mu}{32\pi r^3}
	\left(\frac{m_2}{m_1}\right)
	\vec\ell
$
&
\ccell
$\tablestyle
	\frac{m \mu}{128 \pi^2 r^6}
	\left(\frac{m_2}{m_1}\right)
	\vec\ell
$
\\
$ \Delta\vec\Omega_1^{(n)}\scount{ss} $
&
$0$
&
\ccell
$\tablestyle
	-\frac{\mu}{128\pi^2r^6}
	\bigg(
		\vec v \times (\vec v \times \vec S_2)
		-
		3 (\vec n \cdot \vec v)
		(\vec S_2 \times \vec n)\times \vec v
		+
		\frac{3(\vec\ell \cdot \vec S_2)\vec\ell}
		{r^2}
	\bigg)
$
\\
\hline
\end{tabular}
\caption{Leading-order equations of motion for a binary system of spinning point-like bodies coupled conformally and disformally to a light scalar field. The expressions highlighted in blue are novel to this work. Expressions for the scalar-induced corrections ${\Delta \vec a \equiv \Delta\vec a_1 - \Delta\vec a_2}$ to the relative acceleration of the binary are catalogued in the first three rows, while the remaining two give the scalar-induced corrections~$\Delta\vec\Omega_1$ to the precession rate of the first body's spin vector. One obtains $\Delta\vec\Omega_2$ by simply interchanging the labels~${1\leftrightarrow 2}$, and note that the vectors $\{\vec n, \, \vec r, \, \vec v, \, \vec a \}$ swap sign under this interchange. All of the symbols used in this work are defined in sections~\ref{sec:intro_summary} and~\ref{sec:intro_conventions}.}
\label{table:summary}
\end{table}
\egroup

In principle, there are thus 10 different contributions that need to be calculated, but expressions for six of these are already known. In the conformal sector, both $\Delta\vec a_A^{(0)}$ and $\Delta\vec\Omega_A^{(0)}$ can be obtained by simply mapping the ${\dis/\M^4 \to 0}$ limit of our scalar-tensor theory onto the parametrised post-Newtonian (PPN) formalism~\cite{Will:2018},%
\footnote{Specifically, the sums of the general relativistic and conformal parts follow from eqs.~(6.79), (6.113), and (6.121) of ref.~\cite{Will:2018} after reinstating factors of the gravitational constant~$\GN$, rescaling ${ \GN \to \GN (1 + \con^2/2) }$, and setting the PPN parameter ${ \gamma_\text{PPN} = (2 - \con^2)/(2 + \con^2) }$, with all other PPN parameters set to their general relativistic values. Additionally, the gauge parameter~$\lambda$ in eq.~(6.113) is set equal to $1$ to impose the covariant spin supplementary condition (see section~\ref{sec:gr} for details).}
 while in the disformal sector, the leading-order expression for $\Delta\vec a_A^{(1)}\scount{o}$ (but limited to two bodies) can be obtained from the Lagrangian recently derived in ref.~\cite{Brax:2019tcy}. To complete this picture, in section~\ref{sec:disformal} we derive novel expressions for the disformal spin-orbit and spin-spin accelerations, $\Delta\vec a_A^{(1)}\scount{so}$ and $\Delta\vec a_A^{(1)}\scount{ss}$, and the disformal spin-orbit and spin-spin precession rates, $\Delta\vec\Omega_A^{(1)}\scount{so}$ and $\Delta\vec\Omega_A^{(1)}\scount{ss}$. Our results are valid for any number~$N$ of point-like bodies, but for ease of reference, explicit expressions for the especially interesting case of binary systems (${N=2}$) are presented here in table~\ref{table:summary}.
\looseness=-1

\begin{figure}
\centering\includegraphics[width=\textwidth]{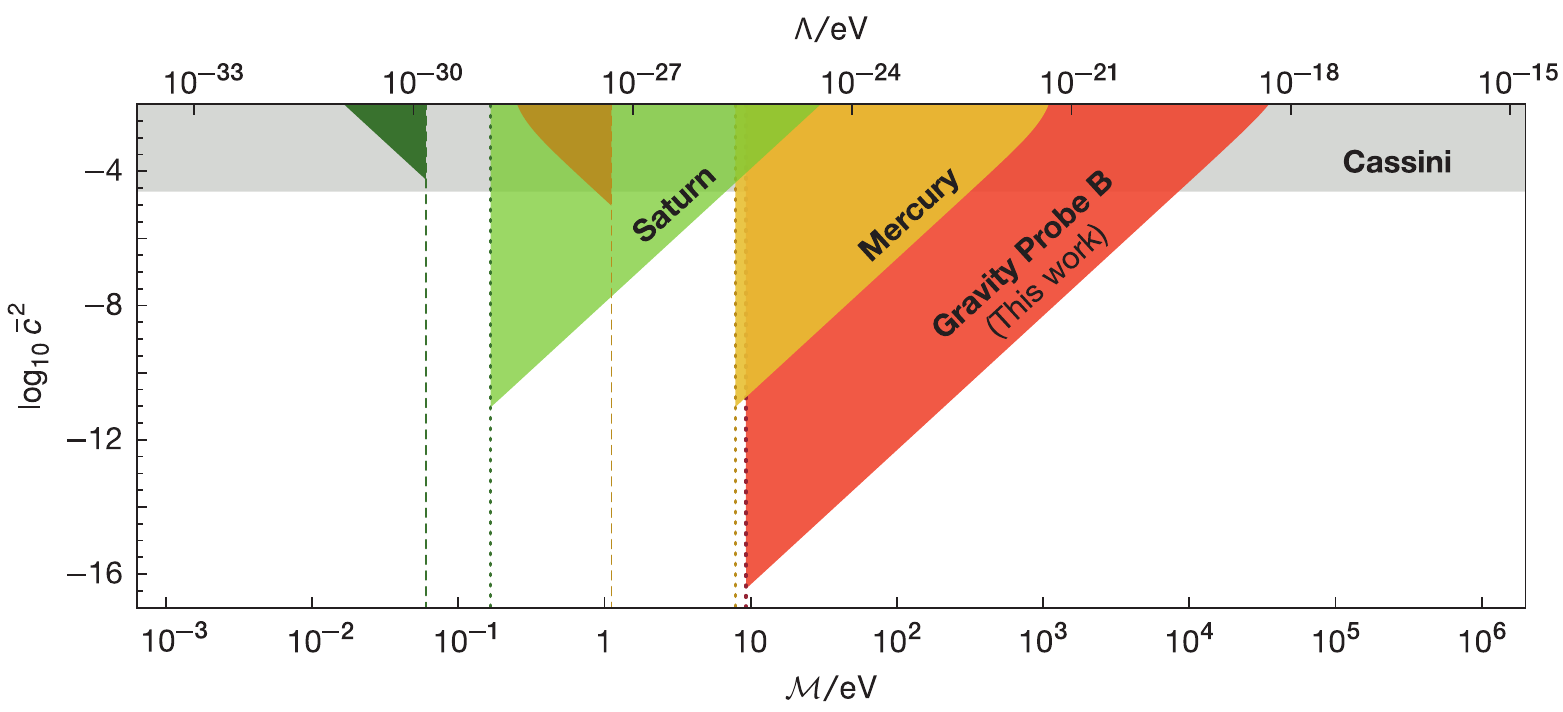}
\caption{Constraints on the $\con^2$--$\M$ plane (with ${\dis=1}$) due to a selection of Solar System tests. Shaded regions are excluded at the 95\% confidence level. We note that for each system, there is an associated threshold value of~$\M$ (indicated by a dotted line) below which a perturbative expansion in powers of $1/\M^4$ breaks down. As the calculations in this paper are predicated on the validity of this ladder expansion, we use spin-precession measurements from the Gravity Probe~B experiment to establish constraints only in the region above this threshold (to the right of the red dotted line). This newly excluded region is shown alongside existing constraints from the perihelion precessions of Mercury and Saturn, for which the light yellow and light green regions rely on similar perturbative calculations~\cite{Brax:2018bow, Davis:2019ltc}. Additionally, for sufficiently small values of~$\M$, further constraints (i.e., the dark yellow and dark green regions to the left of the corresponding dashed lines) can be established due to recent work on resumming the ladder expansion~\cite{Davis:2019ltc}. Also drawn on this figure is the upper bound on $\con^2$ due to the Cassini spacecraft~\cite{Bertotti:2003rm}, but note that this is a naive constraint that does not take potential disformal effects into account. Along the top axis, we show how the disformal scale~$\M$ is related to the scale~${\Lambda \equiv \M^2/\Mpl}$; another parametrisation often used in the literature.}
\label{fig:constraints_M}
\end{figure}

Armed with these results, in section~\ref{sec:GPB} we confront our predictions with spin-precession measurements from the Gravity Probe~B experiment~\cite{Everitt:2011hp, Everitt_2015} to establish new observational constraints on the disformal coupling. The key plot is shown here in figure~\ref{fig:constraints_M}, where the region of parameter space newly excluded by Gravity Probe~B is shaded in red.
This comes from translating the good agreement between the experimental results and the theoretical predictions from general relativity into an upper bound on the size of the scalar-induced correction~$\Delta\vec\Omega_A$, cf.~\eqref{eqn:GBP_bound_NS}, which simplifies to
\begin{equation}
\ceq{
	\con^2
	\left( \frac{9.2~\text{eV}}{|\dis|^{-1/4} \M} \right)^4
	<
	4.1 \times 10^{-17}
}
\label{eqn:intro_bound_GPB}
\end{equation}
for values of $\con^2$~($< 2.5 \times 10^{-5}$) not already ruled out by the Cassini spacecraft~\cite{Bertotti:2003rm}. It is worth emphasising that this constraint is predicated on the validity of the ladder expansion in \eqref{eqn:intro_expansion_ladder}, which holds only when ${|\dis|^{-1/4}\M \gg E_c}$, where $E_c$ is some characteristic energy scale of the system. For Gravity Probe~B, we will argue in later sections that $E_c \sim 9.2~\text{eV}$; hence, we may confidently rule out only the region of parameter space that violates \eqref{eqn:intro_bound_GPB} but satisfies
\begin{equation}
	|\dis|^{-1/4}\M \gg 9.2~\text{eV}.
\label{eqn:intro_bound_GPB_ladder_cutoff}
\end{equation}

Presented alongside this newly excluded region are existing constraints due to the perihelion precession of Mercury and Saturn~\cite{Brax:2018bow, Davis:2019ltc}. For Mercury, the constraints and their corresponding regimes of validity are
\begin{subequations}
\label{eqn:intro_bound_Mercury}
\begin{alignat}{3}
	\con^2\bigg( \frac{7.8~\text{eV}}{|\dis|^{-1/4}\M} \bigg)^4
	&<
	1.4 \times 10^{-11}
	&&\qquad
	(\text{valid when } |\dis|^{-1/4}\M \gg 7.8~\text{eV}),
	\label{eqn:intro_bound_Mercury_expansion}
\allowdisplaybreaks\\
	\con^2\bigg( \frac{|\dis|^{-1/4}\M}{1.1~\text{eV}} \bigg)^4
	&<
	1.4 \times 10^{-5}
	&&\qquad
	(\text{valid when } |\dis|^{-1/4}\M \ll 1.1~\text{eV}),
	\hspace{10.5pt}
	\label{eqn:intro_bound_Mercury_resum}
\end{alignat}
\end{subequations}
while for Saturn, we have
\begin{subequations}
\label{eqn:intro_bound_Saturn}
\begin{alignat}{3}
	\con^2\bigg( \frac{0.17~\text{eV}}{|\dis|^{-1/4}\M} \bigg)^4
	&<
	1.4 \times 10^{-11}
	&&\qquad
	(\text{valid when } |\dis|^{-1/4}\M \gg 0.17~\text{eV}),
	\label{eqn:intro_bound_Saturn_expansion}
\allowdisplaybreaks\\
	\con^2\bigg( \frac{|\dis|^{-1/4}\M}{0.060~\text{eV}} \bigg)^4
	&<
	8.6 \times 10^{-5}
	&&\qquad
	(\text{valid when } |\dis|^{-1/4}\M \ll 0.060~\text{eV}).
	\label{eqn:intro_bound_Saturn_resum}
\end{alignat}
\end{subequations}
The upper bounds in \eqref{eqn:intro_bound_Mercury_expansion} and \eqref{eqn:intro_bound_Saturn_expansion} follow from confronting eq.~(5.7) of ref.~\cite{Davis:2019ltc}, which is valid to first order in $\dis/\M^4$, with the observational bounds in ref.~\cite{10.1093/mnras/stt695}, and result in the  light yellow and light green regions in figure~\ref{fig:constraints_M}, respectively. Meanwhile, the bounds in \eqref{eqn:intro_bound_Mercury_resum} and \eqref{eqn:intro_bound_Saturn_resum} follow from eq.~(5.11) of ref.~\cite{Davis:2019ltc}, where it was shown that for sufficiently small values of~$|\dis|^{-1/4}\M$, it is possible to resum the ladder expansion and then reorganise it as a new perturbative series in powers of~$\M^4/\dis$. This leads to the dark yellow and dark green regions in figure~\ref{fig:constraints_M}. 
Although, for both Mercury and Saturn, these dark and light regions do not overlap,%
\footnote{We expect that a fully nonperturbative method will be required to fill in the remaining gap between the corresponding dashed and dotted lines. Such a result will also likely smooth out the straight edges of the shaded regions where they meet these vertical lines, since the first-order perturbative calculations used to derive \eqref{eqn:intro_bound_GPB}, \eqref{eqn:intro_bound_Mercury}, and \eqref{eqn:intro_bound_Saturn} become unreliable as we approach the threshold where perturbation theory breaks down. However, as figure~\ref{fig:constraints_M} is drawn on a logarithmic scale, the region in which we incur such errors is likely to be small, and so we shall assume that this figure is still reasonably accurate.}
any reasonable interpolation between the two will tell us that the orbital motion of a binary is most strongly influenced by (spin-independent) disformal effects when the coupling scale~$|\dis|^{-1/4}\M$ is comparable to the system's characteristic energy scale~$E_c$. The same is likely to be true also for the spin-\emph{dependent} disformal effects, but a proper check would require a handle on resumming the spin-dependent terms in the ladder expansion, which we leave for the future.

For now, it is interesting to ask how the different constraints, as they appear in figure~\ref{fig:constraints_M}, fare against one another. Although each one is most effective at probing a different region of parameter space, one way in which we can meaningfully compare their relative constraining power is to ask how far each shaded region extends down the vertical axis. By this measure, Gravity Probe~B places the strongest constraints (we get an improvement of more than 5~orders of magnitude over the constraints set by Mercury and Saturn's orbits), and so our results position spin effects as a powerful probe of disformal interactions. In section~\ref{sec:disc}, we comment briefly on several avenues for taking these promising results even further.
\looseness=-1

\subsection{Conventions}
\label{sec:intro_conventions}

We work in ${3 + 1}$ spacetime dimensions with metric signature $(-,+,+,+)$ and employ units in which ${\hbar=c=1}$ throughout, with the reduced Planck mass given by $\Mpl \equiv (8\pi \GN)^{-1/2}$.
Spacetime indices are labelled by Greek letters $\{ \mu, \nu , \,\dots \}$ that run from 0 to 3, while spatial indices are labelled  by lowercase Latin letters $\{ i, j , \,\dots \}$ running from 1~to~3. In the weak-field limit, we will always work in Cartesian bases for which there is no distinction between a raised or lowered spatial index, and so will often position these indices wherever is most convenient. The fully antisymmetric Levi--Civita symbol is defined with the sign convention ${\epsilon_{0123} = +1}$, and the corresponding tensor density is ${ \varepsilon_{\mu\nu\rho\sigma} = \sqrt{-g}\epsilon_{\mu\nu\rho\sigma} }$.

Uppercase Latin letters $\{ A , B , \,\dots \}$ running from 1 to $N$ label the different bodies in our system. Unlike the spacetime and spatial indices, these labels are \emph{not} implicitly summed over. Both the spin vector $S_A^i$ and the antisymmetric spin tensor ${ S_A^{ij} \equiv \epsilon^{ijk} S_A^k }$ will be used interchangeably (${ \epsilon^{ijk} \equiv -\epsilon^{0ijk} }$), but for any other vector $\vec X$, we adopt conventional multi-index notation for writing the product of $n$ vectors as ${ X^{i_1 \cdots\, i_n} \equiv X^{i_1} \cdots X^{i_n} }$. Angled brackets around indices denote the symmetric, trace-free projection; e.g., $X^\avg{ij} = X^i X^j - \vec X^2 \delta^{ij}/3$.

At a given coordinate time~$t$, we denote the displacement between  the bodies $A$ and $B$ by ${ \vec r_{AB} = \vec x_A(t) - \vec x_B(t) }$, with corresponding unit vector ${ \vec n_{AB} = \vec r_{AB}/r_{AB} }$, where ${ r_{AB} = |\vec r_{AB}| }$. Similarly, we write ${ \vec r_A = \vec x - \vec x_A(t) }$ as shorthand for the displacement from body~$A$ to some general point~$\vec x$. The 3-velocities and 3-accelerations of the bodies are ${\vec v_A = \dx\vec x_A/\dx t}$ and ${\vec a_A = \dx\vec v_A/\dx t}$, respectively; from which we can define the relative velocities ${\vec v_{AB} = \vec v_A - \vec v_B}$ and the relative accelerations ${\vec a_{AB} = \vec a_A - \vec a_B}$.

For binary systems with just ${N=2}$ bodies, we abbreviate this notation further by writing ${ \vec r = \vec r_{12} }$, ${ \vec n = \vec n_{12} }$, ${ \vec v = \vec v_{12} }$, and ${ \vec a = \vec a_{12} }$. For two bodies with masses $m_1$ and $m_2$, we use ${ m = m_1 + m_2 }$ and ${ \mu = m_1 m_2/m }$ to denote their total and reduced masses, respectively. Lastly, it also proves convenient to define ${\vec\ell = \vec r \times \vec v}$ as the specific angular momentum of the effective-one-body system, and to introduce two particular linear combinations of the spin vectors:
${ \vec S_\pm = m_2 \hat{\vec S}_1 \pm m_1\hat{\vec S}_2 }$, where ${ \hat{\vec S}_A = \vec S_A/m_A }$.

\section{Spinning bodies in general relativity}
\label{sec:gr}

As a warm up to the scalar-tensor case, we begin in this section with a review of spinning point particles in general relativity. We take as our starting point the Mathisson--Papapetrou--Dixon equations \cite{Mathisson:1937zz, Papapetrou:1951pa, Dixon:1970zza, Dixon:1970zz, Dixon:1974},%
\footnote{
This corresponds to a minimal coupling between the spinning point particle and the spacetime curvature\,---\,see, e.g., refs.~\cite{Deriglazov:2017jub, Deriglazov:2018vwa} for the effect of possible nonminimal couplings.}
\begin{equation}
\label{eqn:gr_MPD} 
	\dot p_\mu
	=
	- \frac{1}{2} R_{\mu\nu\rho\sigma} u^\nu S^{\rho\sigma},
	\qquad
	\dot{S}^{\mu\nu}
	=
	2p^{[\mu} u^{\nu]},
\end{equation}
which govern the evolution of a test particle with 4-velocity~$u^\mu$, 4-momentum~$p^\mu$, and spin-angular-momentum tensor~$S^{\mu\nu}$ that is travelling on some background spacetime with metric~$g_{\mu\nu}$. (The overdot denotes a covariant derivative along the worldline; i.e., $\,\dot{\vph{a}}\, \equiv u^\mu\nabla_\mu$.)

The spin tensor~$S^{\mu\nu}$ is antisymmetric by construction, and so the set $\{ u^\mu, p^\mu, S^{\mu\nu} \}$ has a total of 14 degrees of freedom, but only six of these are physical. Three coordinates are needed to specify the position of this particle in space and three angles are needed to specify its orientation relative to some space-fixed frame; the other eight are gauge degrees of freedom whose inclusion is what facilitates a generally covariant description of this particle. To solve \eqref{eqn:gr_MPD}, however, requires that we fix a gauge.

We first remove the three redundant degrees of freedom in~$S^{\mu\nu}$, which correspond to a freedom of choice in how one defines the particle's effective centre-of-energy coordinate, by imposing a spin supplementary condition (SSC)~\cite{Hanson:1974qy, Pryce:1948pf}. We have found it most convenient to work with the covariant SSC,
\begin{equation}
	S^{\mu\nu} p_\nu \approx 0,
\label{eqn:gr_SSC_p}
\end{equation}
throughout this paper. Requiring that this condition holds along the entirety of the particle's worldline removes another four gauge degrees of freedom by establishing the relation~\cite{Foffa:2013qca, Porto:2016pyg}
\begin{equation}
	p^\mu
	=
	\frac{1}{\sqrt{-u^2}}
	\left(
		m u^\mu
		-
		\frac{1}{2m} S^{\mu\nu} R_{\nu\lambda\rho\sigma}
		u^\lambda S^{\rho\sigma}
		+
		\mathcal O(R^2S^4)
	\right),
\label{eqn:gr_p(u)}
\end{equation}
where the constant ${ m = \sqrt{- p_\mu p^\mu}}$ is the rest mass of the particle. Crucially, note that in contrast to a nonspinning body, a spinning one need not have $p^\mu$ parallel to $u^\mu$ when the spacetime is curved. The last remaining gauge degree of freedom is the most familiar, and is associated with reparametrisation invariance of the worldline. We remove it by enforcing the normalisation condition
\begin{equation}
	u_\mu u^\mu = -1.
\label{eqn:u^2}
\end{equation}

Now substituting \eqref{eqn:gr_p(u)} back into \eqref{eqn:gr_MPD}  gives us equations of motion for just $u^\mu$ and $S^{\mu\nu}$,
\begin{subequations}
\label{eqn:gr_eom} 
\begin{align}
	\dot u_\mu &=
	-\frac{1}{2m} R_{\mu\nu\rho\sigma}
	u^\nu S^{\rho\sigma}
	+
	\cdots,
	\label{eqn:gr_eom_u} 
	\\
	\dot S^{\mu\nu} &= 0 + \cdots,
	\label{eqn:gr_eom_S} 
\end{align}
\end{subequations}
which we have to solve alongside the constraints in~\eqref{eqn:gr_SSC_p} and \eqref{eqn:u^2}. The ellipses above allude to terms that are of quadratic order or higher in the spin, which we will neglect. In fact, as we are working up to linear order in the spin, the SSC simplifies to
\begin{equation}
	S^{\mu\nu} u_\nu \approx 0.
\label{eqn:gr_SSC}
\end{equation}

\subsection{Acceleration}

We shall now discuss how to solve \eqref{eqn:gr_eom_u} and \eqref{eqn:gr_eom_S} in turn, assuming henceforth that the background metric is only weakly curved, such that we may write ${g_{\mu\nu} = \eta_{\mu\nu} + h_{\mu\nu}}$. Expanding \eqref{eqn:gr_eom_u} to linear order in $h_{\mu\nu}$ gives
\begin{equation}
	u^\nu \partial_\nu u^\mu
	=
	\eta^{\mu\nu}
	\left(
		\frac{1}{2}\partial_\nu h_{\rho\sigma}
		-
		\partial_{\rho} h_{\sigma\nu}
	\right)
	u^\rho u^\sigma
	+
	\eta^{\mu\lambda}\hat S^{\rho\sigma} u^{\nu}
	\partial_\rho \partial_{[\lambda} h_{\nu]\sigma}
	+
	\mathcal O(h^2),
\label{eqn:gr_a_tau}
\end{equation}
where we define ${\hat S^{\rho\sigma} = S^{\rho\sigma}/m}$ for brevity, and it is to be understood that $h_{\mu\nu}$ and its derivatives are being evaluated along the worldline. Since ${ u^\mu \coloneq dx^\mu/d\tau }$, the result is a set of ordinary differential equations for the worldline components ${ x^\mu \equiv x^\mu(\tau) }$. 

It proves more convenient, however, to parametrise this worldline with respect to the coordinate time~$t$~(${\equiv x^0}$), rather than the proper time~$\tau$. Moving from one to the other is achieved by defining a new velocity vector ${ v^\mu \coloneq \dx x^\mu/\dx t = u^\mu/u^0 }$, where the normalisation~factor
\begin{equation}
	u^0 = [1 - \vec v^2 - h_{\mu\nu} v^\mu v^{\nu \vph{\mu}} + \mathcal O(h^2)]^{-1/2}
\label{eqn:gr_u0}
\end{equation}
follows from the constraint in \eqref{eqn:u^2}. Written in terms of the 3-acceleration ${a^i \coloneq \dx v^i/\dx t}$ and the usual Lorentz factor ${ \gamma = 1/\sqrt{1-\vec v^2} }$, \eqref{eqn:gr_a_tau} now reads
\begin{align}
	[ \delta_{ij} + \gamma^2 v_i v_j + \mathcal O(h)]\, a^j
	&=
	\frac{1}{2}
	(
		\partial_i h_{\mu\nu}
		-
		2\partial_{\mu} h_{\nu i}
		-
		\gamma^2 v_i  \partial_0 h_{\mu\nu}
	)
	v^\mu v^\nu
	\nonumber\\&\quad
	+
	\gamma^{-1} \hat S^{\mu\nu} v^{\lambda}
	\partial_\mu \partial_{[i} h_{\lambda]\nu}
	+
	\mathcal O(h^2).
\label{eqn:gr_a_1PM}
\end{align}

This differential equation is still challenging to solve, but a perturbative solution can be obtained when the 3-velocity ${v^i \ll 1 }$, in which case the projection matrix ${\delta_{ij} + \gamma^2 v_i v_j}$ on the lhs can be inverted order-by-order in~$v$. To reproduce the Newtonian limit correctly, the leading-order solution to this equation must~be
\begin{equation}
	a_i = \frac{1}{2} \partial_i h_{00} + \cdots,
\label{eqn:gr_a_Newtonian}
\end{equation}
where one identifies $-h_{00}/2$ as the Newtonian gravitational potential. If the orbit is bound, then ${a^i \sim v^2/r}$ on dimensional grounds, and so the above equation tells us that ${h_{00} \sim v^2}$. One can further show that ${ h_{ij} \sim v^2 }$ also, but ${ h_{0i} \sim v^3 }$~\cite{MTW}.

\subsection{Spin precession}

Consider now the evolution of the particle's spin, governed by \eqref{eqn:gr_eom_S}. Rather than work with the antisymmetric tensor $S^{\mu\nu}$ directly, it is useful to introduce the Pauli--Lubanski spin vector
\begin{equation}
	S_\mu
	\coloneq
	-\frac{1}{2m} \varepsilon_{\mu\nu\rho\sigma} p^\nu S^{\rho\sigma}
	=
	-\frac{1}{2} \varepsilon_{\mu\nu\rho\sigma} u^\nu S^{\rho\sigma} + \cdots,
\label{eqn:gr_pauli_lubanski_vector}
\end{equation}
which contains only the three physical degrees of freedom in $S^{\mu\nu}$ that remain after we impose the covariant SSC~\eqref{eqn:gr_SSC_p}.
Equations~\eqref{eqn:gr_p(u)} and \eqref{eqn:gr_eom} then tell us that ${ \dot S_\mu = 0 }$; i.e., that the particle's spin vector is parallel transported along its worldline. That being said, what can be measured in practice are the components ${S^\vph{\mu}_{\hat\alpha} = e_{\hat\alpha}^\mu S^\vph{\mu}_\mu}$ of this vector projected onto a frame that is comoving with the particle, and if this comoving frame is not also parallel transported along the worldline, then the components $S_{\hat\alpha}$ will generally be seen to precess.\looseness=-1

Erecting an orthonormal tetrad $\{ e^\mu_{\hat\alpha} \}$ with local Lorentz indices $\{\hat\alpha, \hat\beta,\,\dots \}$ on this worldline defines for us the Jacobian ${ e^\mu_{\hat\alpha} \equiv \partial x^\mu/\partial y^{\hat\alpha} }$ for transforming between the global $x^\mu$ coordinates that we have been using thus far (corresponding to an inertial observer at rest at infinity) and the local coordinates $y^{\hat\alpha}$ for this comoving frame, which we shall require to satisfy two properties. First, the metric expressed in these comoving coordinates should be that of Minkowski at all points along the worldline, and second, the spatial coordinates~$y^{\hat\imath}$ should not be rotating with respect to the $x^\mu$~frame. One effects such a transformation by simply rescaling the coordinates and then performing a pure Lorentz boost to align the timelike vector~$e^\mu_{\hat 0}$ along the direction of the 4-velocity~$u^\mu$; see, e.g., Chapter~39.10 of ref.~\cite{MTW} for more details. The end result is that ${ e^\mu_{\hat 0} = u^\mu_\vph{\hat 0} }$ exactly, while the three spatial vectors $e^\mu_{\hat\imath}$ are given by\looseness=-1
\begin{subequations}
\label{eqn:gr_comoving_tetrad}
\begin{align}
	e^0_{\hat\imath}
	&=
	\left( 1 + \frac{1}{2} v^2 + h_{00} \right) v_i
	+
	\frac{1}{2} h_{ij}v^j + h_{0i}
	+
	\mathcal O(v^5),
	\\
	e^i_{\hat\jmath}
	&=
	\delta^i_j
	+
	\frac{1}{2} (v^i v_j - h^i{}_j)
	+
	\mathcal O(v^4).
\end{align}
\end{subequations}

Simple use of the chain rule now tells us that
\begin{equation}
	\d{}{\tau} S_{\hat\alpha}
	=
	u^\nu \nabla_\nu ( S_\mu e^\mu_{\hat\alpha})
	=
	S_\mu (u^\nu \nabla_\nu e^\mu_{\hat\alpha}),
\end{equation}
and since the vectors $e^\mu_{\hat\alpha}$ are not being parallel transported along the worldline, we may write generically that
\begin{equation}
	u^\nu \nabla_\nu e^\mu_{\hat\alpha}
	=
	- \Omega^\vph{\mu}_{\hat\alpha}{}^{\hat\beta} e^\mu_{\hat\beta}.
\end{equation}
Orthonormality of this tetrad implies that the inner products ${ g_{\mu\nu} e^\mu_{\hat\alpha} e^{\nu\vph{\mu}}_{\hat\beta} = \eta_{\hat\alpha\hat\beta} }$ are constant; hence, $\Omega_{\hat\alpha\hat\beta}$ is an antisymmetric tensor. Notice, furthermore, that the definition in \eqref{eqn:gr_pauli_lubanski_vector} implies that ${S_{\hat 0} = u^\mu S_\mu = 0}$, while \eqref{eqn:gr_eom_u} tells us that $\Omega_{\hat 0 \hat\beta} = 0$ up to $\mathcal O(S)$ corrections, which we neglect. Putting all of this together, it follows that only the spin vector's spatial components~$S_{\hat\imath}$ in this comoving frame are nontrivial, and that they evolve according to the precession equation
\begin{equation}
	\d{}{\tau} S_{\hat\imath}
	=
	-\Omega_{\hat\imath\hat\jmath} S^{\hat\jmath},
	\qquad
	\Omega_{\hat\imath\hat\jmath}
	=
	g_{\mu\nu} e^\mu_{\hat\imath} u^\rho \nabla_\rho e^\nu_{\hat\jmath}.
\label{eqn:gr_eom_proper_spin}
\end{equation}
This equation can be rewritten more conveniently in vectorial form,
\begin{equation}
	\d{S_{\hat\imath}}{\tau} = (\vec\Omega \times \vec S)_{\hat\imath},
\label{eqn:gr_eom_proper_spin_vectorial}
\end{equation}
by defining the angular velocity vector ${ \Omega^{\hat\imath} = \epsilon^{\hat\imath\hat\jmath\hat k}\Omega_{\hat\jmath\hat k}/2 }$. If desired, the evolution of $S_{\hat\imath}$ with respect to coordinate time can then be obtained by using the relation ${\dx t/\dx\tau = u^0}$, and note that ${u^0 = 1 + \mathcal O(v^2)}$.

All that remains is to obtain an explicit expression for $\vec \Omega$. Substituting \eqref{eqn:gr_comoving_tetrad} into \eqref{eqn:gr_eom_proper_spin} and identifying ${ e^\mu_{\hat 0} = u^0 v^\mu }$, we find
\begin{equation}
	\Omega^{\hat\imath}
	=
	\frac{1}{4}\epsilon^{ijk}
	(
		v^j \partial_k h_{00}
		-
		2 v^\mu\partial_j h_{k\mu}
	)
\label{eqn:gr_Omega_vector} 
\end{equation}
to leading PN order, after also using \eqref{eqn:gr_a_Newtonian}. One final comment at this stage: notice that the components~$S_i$ and $S_{\hat\jmath}$ as measured in the two frames are related by ${ S_i = e_i^{\hat\jmath} S_{\hat\jmath} = S_{\hat\imath} + \mathcal O(v^2) }$, where ${ e_i^{\hat\jmath} = g_{i\mu} \eta^{\hat\jmath\hat\alpha} e^\mu_{\hat\alpha} }$; hence, drawing a distinction between the two is thus unnecessary at leading PN order when they appear on the rhs of \eqref{eqn:gr_eom_proper_spin_vectorial}, or when they appear in \eqref{eqn:gr_a_1PM}. However, it \emph{is} necessary to distinguish between $\dx S_{\hat\jmath}/\dx t$ and $\dx S_i/\dx t$, since these equations of motion differ by terms associated with the time evolution of $e_i^{\hat\jmath}$.

\section{Spinning bodies in disformal scalar-tensor theories}
\label{sec:disformal}

Our goal now is to extend the preceding results to disformal scalar-tensor theories. We shall restrict our attention to metric theories that are described by the general action
\begin{equation}
	S = S_\text{ST}[g,\phi]
	+
	S_m[\tilde g,\Psi].
\label{eqn:stt_action}
\end{equation}
The kinetic terms for, and the interactions between, the Einstein-frame metric~$g_{\mu\nu}$ and scalar~$\phi$ are encoded in the scalar-tensor action $S_\text{ST}$. These fields are assumed to couple universally to matter (collectively denoted by~$\Psi$) via the effective Jordan-frame metric~\cite{Bekenstein:1992pj}
\begin{equation}
	\tilde g_{\mu\nu}
	=
	C(\phi ,X) g_{\mu\nu}
	+
	D(\phi,X) \partial_\mu\phi\,\partial_\nu\phi
	\qquad
	(X \equiv g^{\mu\nu}\partial_\mu\phi\,\partial_\nu\phi).
\label{eqn:stt_Jordan_frame}
\end{equation}

Naturally, there is a good amount of freedom when it comes to choosing the conformal and disformal factors, $C(\phi,X)$ and $D(\phi,X)$, but from the perspective of a low-energy effective field theory, their most relevant features can be captured with just a few coefficients. To leading order in an expansion in the small ratios $\phi/\Mpl$ and~$\partial\phi/\M^2$, we write
\begin{equation}
	C(\phi,X) = 1 + \frac{\con \hspace{1pt} \phi}{\Mpl} + \cdots,
	\qquad
	D(\phi,X) = \frac{\dis}{\M^4} + \cdots,
\label{eqn:stt_EFT_expansion}
\end{equation}
where we have chosen a normalisation such that ${ \tilde g_{\mu\nu} \to g_{\mu\nu} }$ when ${\phi\to 0}$. Note that the dimensionless parameter $\con$ controls the strength of the conformal coupling relative to gravity, but the strength of the disformal coupling is set by the combination $\dis/\M^4$, and so we are always free to rescale $\M$ such that ${|\dis|=1}$. The parameter $\dis$ cannot be removed entirely, however, as it is needed to keep track of the overall sign of this interaction. (To~give rise to order-one deviations from general relativity on cosmological scales, one typically sets the scale ${\Lambda \equiv \M^2/\Mpl}$ equal to the Hubble constant~$H_0$, but we shall here impose no prior restrictions on~$\M$ to keep our analysis as general as possible.)

Returning to the scalar-tensor sector, we shall further assume for simplicity that any nonlinear and/or nonminimal interactions in~$S_\text{ST}$ are subleading on the scales that we are interested in, and so will take as our fiducial model general relativity plus a massless scalar:
\begin{equation}
	S_\text{ST}[g,\phi]
	=
	\frac{1}{2}\int\dx^4x\sqrt{-g}
	\left(
		\Mpl^2 R - g^{\mu\nu}\partial_\mu\phi\,\partial_\nu\phi
		+
		\cdots
	\right).
\end{equation}

The key question now is: how does the addition of this scalar field affect the evolution of a spinning body? In general relativity, the Mathisson--Papapetrou--Dixon equations in~\eqref{eqn:gr_MPD} can be shown to be a direct consequence of the conservation of energy and momentum, ${\nabla_\mu T^{\mu\nu} = 0 }$~\cite{Mino:1995fm, Porto:2005ac}. In our scalar-tensor theory, energy and momentum are conserved with respect to the Jordan-frame metric~$\tilde g_{\mu\nu}$, and so it follows that the evolution of a spinning body in this theory is also governed by \eqref{eqn:gr_MPD}, except with the substitution ${g_{\mu\nu} \to \tilde g_{\mu\nu}}$, or~equivalently, $h_{\mu\nu} \to \tilde h_{\mu\nu}$ (${\equiv \tilde g_{\mu\nu} - \eta_{\mu\nu}}$) when working in the weak-field limit. Making this substitution in~\eqref{eqn:gr_a_1PM} and \eqref{eqn:gr_eom_proper_spin}, and also generalising to the case of $N$~spinning point-like bodies, we see that the 3-acceleration of the $A$th body is given by
\begin{subequations}
\label{eqn:disformal_master}
\begin{align}
	[ \delta^{ij} + \gamma_A^{2\vph{j}} v_A^{i\vph{j}} v_A^j + \mathcal O(\tilde h)]\, a_A^j
	&=
	\frac{1}{2}
	(
		\partial_i \tilde h_{\mu\nu}
		-
		2\partial_{\mu} \tilde h_{\nu i}
		-
		\gamma_A^2 v_A^i  \partial_0 \tilde h_{\mu\nu}
	)
	v_A^\mu v_A^{\nu \vph{\mu}}
	\nonumber\\&\quad
	+
	\gamma_A^{-1} \hat S_A^{\mu\nu} v_A^{\lambda}
	\partial_\mu \partial_{[i} \tilde h_{\lambda]\nu}
	+
	\mathcal O(\tilde h^2)
	\label{eqn:disformal_master_acc}
\end{align}
in this scalar-tensor theory, while its precession rate is
\begin{equation}
	\Omega_A^{\hat\imath}
	=
	\frac{1}{4}\epsilon^{ijk}
	(
		v_A^j \partial_k \tilde h_{00}
		-
		2 v_A^\mu\partial_j \tilde h_{k\mu}
	).
	\label{eqn:disformal_master_spin}
\end{equation}
\end{subequations}

To proceed, we require knowledge of the metric perturbation~$\tilde h_{\mu\nu}$ sourced by these $N$ bodies, which we derive in section~\ref{sec:disformal_profile}. Then, in section~\ref{sec:disformal_acc}, we substitute this solution back into \eqref{eqn:disformal_master_acc} to obtain an explicit expression for the 3-acceleration. We do the same for the precession rate~\eqref{eqn:disformal_master_spin} in section~\ref{sec:disformal_spin}.

\subsection{Field profiles sourced by \emph{N} spinning bodies}
\label{sec:disformal_profile}

Before launching into detailed calculations, it is instructive to first enumerate the different scales in this problem. We are interested in a system of $N$ spinning bodies evolving on an otherwise flat spacetime, which we shall assume is only weakly curved by the presence of these bodies. All $N$ of them are assumed to be widely separated such that each one can be treated as approximately point-like, and each one is further assumed to be travelling nonrelativistically. Purely for the purposes of power counting, let us suppose that all of these separations $\vec r_{AB}$ are on the order of some characteristic length scale~$r$, and similarly let all of the 3-velocities $\vec v_A$ be on the order of some typical velocity~$v$. Additionally, let $m$ be some characteristic mass scale related to the masses $m_A$ of these bodies, and let $S$ be some characteristic value for their spins~$\vec S_A$.

\paragraph{Expansion parameters.}
From these, we can identify four dimensionless expansion parameters about which to organise our perturbative solution: 
(1)~the characteristic velocity~$v$ of the bodies,
(2)~the post-Minkowskian parameter ${\epsilon_\text{PM} \sim \GN m/r}$,
(3)~the ladder parameter ${\epsLadder \sim \dis m v^2/(\M^4 r^3)}$, and
(4)~the small ratio ${ \epsSpin \sim S/L }$ between the spin~$S$ of the body and its orbital angular momentum ${ L \sim mrv }$.

While these four parameters control different aspects of the perturbative expansion, we know that for gravitationally bound systems, the first two will be of comparable size due to the virial theorem, ${ v^2 \sim \epsPM }$. This relation holds both in general relativity and also for the scalar-tensor theory considered here, provided only that $\con$ is not too much larger than one%
\footnote{This is guaranteed by the Cassini spacecraft~\cite{Bertotti:2003rm}, which establishes the upper bound ${ \con^2 \leq 2.5 \times 10^{-5} }$.}
and that $\epsLadder$ is suitably small. Thus, while in intermediate steps of the calculation we will power count separately in both $v$ and $\epsPM$, we must make the identification ${\epsPM \sim v^2}$ at the end, and so our solutions are really organised as an expansion in just three parameters: $v$, $\epsLadder$, and $\epsSpin$. \looseness=-1

\paragraph{Field equations.}
Let us now expand the metric and scalar field about flat space by writing
\begin{align}
	g_{\mu\nu}(x) = \eta_{\mu\nu} + h_{\mu\nu}(x),
	\qquad
	\frac{\phi(x)}{\Mpl} = 0 + \varphi(x).
\end{align}
It proves convenient to divide $\phi(x)$ by the Planck mass, such that both $h_{\mu\nu}(x)$ and $\varphi(x)$ are dimensionless. At first order in~$\epsPM$ (i.e., at first post-Minkowskian or 1PM order), the field equations read~\cite{Davis:2019ltc}
\begin{subequations}
\label{eqn:stt_eom_1PM}
\begin{gather}
	\Box
	\bigg(
		h_{\mu\nu}
		-
		\frac{1}{2} h \hspace{1pt}\eta_{\mu\nu}
	\bigg)
	=
	- \frac{2}{\Mpl^2} T_{\mu\nu},
	\label{eqn:stt_eom_1PM_h}
	\allowdisplaybreaks\\
	\Box\varphi
	=
	- \frac{\con}{2 \Mpl^2}T^{\mu\nu}\eta_{\mu\nu}
	+
	\frac{\dis}{\M^4}\partial_\mu (T^{\mu\nu} \partial_\nu\varphi),
	\label{eqn:stt_eom_1PM_varphi}
\end{gather}
\end{subequations}
where ${ \Box \equiv \eta^{\mu\nu}\partial_\mu\partial_\nu }$ is the wave operator on flat space and we have imposed the de~Donder gauge [${\partial^\mu h_{\mu\nu} - (1/2) \partial_\nu h \approx 0}$ with ${h \equiv \eta^{\mu\nu} h_{\mu\nu}}$] in the first line.  On the rhs of these equations, it suffices to evaluate the energy-momentum tensor~$T^{\mu\nu}$ on flat space when working to first order in $\epsPM$,%
\footnote{For this reason, there is no need to distinguish between the energy-momentum tensors defined with respect to the Jordan- and Einstein-frame metrics in this work.}
and so we have that ${ T^{\mu\nu} = \sum_{A=1}^N T^{\mu\nu}_A }$, where~\cite{Dixon1979, Mino:1995fm}
\begin{equation}
	T_A^{\mu\nu}(x)
	=
	\int d\tau_A^\vph{\mu} \, p_A^{(\mu} u_A^{\nu)}
	\,\delta^{(4)}\bm( x - x_A(\tau_A) \bm)
	-
	\partial_\lambda
	\left(
		\int d\tau_A^\vph{(\rho} \, S_A^{\lambda(\mu} u_A^{\nu)}
		\,\delta^{(4)}\bm( x - x_A(\tau_A) \bm)
	\right)
\label{eqn:stt_T}
\end{equation}
is the energy-momentum tensor of the $A$th body. Further note that on flat space, ${p_A^\mu = m_A^\vph{\mu} u_A^\mu}$ exactly once the covariant SSC is imposed; cf.~\eqref{eqn:gr_p(u)}.

A particularly useful feature of these field equations is that they decouple at this post-Minkowskian order; hence, the solution to \eqref{eqn:stt_eom_1PM_h} is exactly the same as in general relativity, and so need not be revisited here.\footnote{See, e.g., eqs.~(9.81), (9.182), and (9.241) of ref.~\cite{PoissonWill} for the result.} Focusing our attention now on \eqref{eqn:stt_eom_1PM_varphi}, we shall assume that the second term on the rhs is suitably small such that we can perform a ladder expansion,
\begin{equation}
\label{eqn:disformal_fields_ladder_expansion}
	\varphi(x) = \frac{\con}{\Mpl^2}
	\bigg(
		\varphi^{(0)}
		+
		\frac{\dis}{\M^4} \varphi^{(1)}
		+
		\mathcal O(\dis^2/\M^8)
	\bigg)
	+
	\mathcal O(1/\Mpl^4),
\end{equation}
where the fields $\varphi^{(0)}$ and $\varphi^{(1)}$ satisfy the equations
\begin{subequations}
\begin{align}
	\Box \varphi^{(0)}
	&=
	- \frac{1}{2} T^{\mu\nu}\eta_{\mu\nu},
	\\
	\Box \varphi^{(1)}
	&=
	\partial_\mu(T^{\mu\nu}\partial_\nu\varphi^{(0)}).
	\label{eqn:disformal_eom_varphi1}
\end{align} 
\end{subequations}
Note that because $\dis/\M^4$ is dimensionful, expanding in powers of this coupling constant is really a proxy for expanding in powers of $\epsLadder$, and likewise expanding in powers of $1/\Mpl^2$ (or~$\GN$) is a proxy for expanding in powers of $\epsPM$.

On each rung~$n$ in this ladder expansion, we may further split ${\varphi^{(n)} = \varphi^{(n)}\scount{o} + \varphi^{(n)}\scount{s}}$, where the first term is the spin-independent part of the field profile, while the second is the spin-dependent part. The solutions for $\varphi^{(0)}\scount{o}$ and $\varphi^{(1)}\scount{o}$ were recently derived in ref.~\cite{Brax:2019tcy} and are reproduced in table~\ref{table:field_profiles} for ease of reference.\footnote{Note that the solution for $\varphi^{(1)}\scount{o}$ in eq.~(2.20) of ref.~\cite{Brax:2019tcy} has a typo. It should read $a_{AB}^i r_{AB}^i$ (as we have it here) instead of $a_{AB}^i n_{AB}^i$.}

\bgroup
\setlength{\tabcolsep}{10pt}
\begin{table}
\small\centering
\begin{tabular}{|ccc|}
\hline
& Conformal $(n=0)$ $\rule[-7pt]{0pt}{20pt}$
& Disformal $(n=1)$ $\rule[-7pt]{0pt}{20pt}$ \\
\hline
$\varphi^{(n)}\scount{o}$
&
$\tablestyle
	- \sum_A \frac{m_A}{8\pi r_A}
$
&
$\tablestyle
	-\sum_A \sum_{B\neq A}
	\frac{m_A m_B}{32 \pi^2 r_A^\vph{3} r_{AB}^3}
	\big(
		a_{AB}^i r_{AB}^i
		-
		3 n_{AB}^\avg{ij} v_{AB}^{ij}
	\big)
$
\\
$\varphi^{(n)}\scount{s}$
&
$0$
&
\ccell
$\tablestyle
	-\sum_A \sum_{B \neq A}
	\frac{3 m_B}{32\pi^2 r_A^2 r^3_{AB} }
	S_A^{ij} n_A^{i \vph{j}} n_{AB}^\avg{jk} v_{AB}^{k \vph{j}}
$
\\
\hline
\end{tabular}
\caption{Scalar-field profiles sourced at leading order by a system of $N$ spinning point-like bodies with masses $m_A$ and spins $S_A$. The expression highlighted in blue is novel to this work. Definitions for all of the symbols can be found in section~\ref{sec:intro_conventions}.}
\label{table:field_profiles}
\end{table}
\egroup%

\paragraph{Deriving the spin-dependent part.}
We are thus left to solve for $\varphi^{(0)}\scount{s}$ and $\varphi^{(1)}\scount{s}$. As the spin-dependent part of the trace $T_A^{\mu\nu}\eta_{\mu\nu}$ is proportional to $S^{\lambda\mu}_A v_A^{\nu \vph{\mu}} \eta_{\mu\nu}$, the former vanishes due to \eqref{eqn:gr_SSC_p}; i.e., ${\varphi^{(0)}\scount{s} = 0}$ at 1PM order when imposing the covariant~SSC.
In contrast, solving \eqref{eqn:disformal_eom_varphi1} via the method of Green functions reveals that $\varphi^{(1)}\scount{s}$ does not vanish at this PM order. Indeed, we find
\begin{align}
	\varphi^{(1)}\scount{s}(x)
	&=
	-\sum_A
	\int d^4x' \, G(x,x')\pd{}{x'^\mu}
	\left(
		T_A^{\mu\nu}\scount{s}(x') \pd{\varphi^{(0)}(x')}{x'^\nu}
	\right)
	\nonumber\\
	&=
	\sum_A \pd{}{x^\mu}
	\int d^4x' \, G(x,x') \pd{}{x'^\lambda}
	\left(
		\int d\tau_A^\vph{(\mu}\, S_A^{\lambda(\mu} u_A^{\nu)}
		\,\delta^{(4)}\bm( x' - x_A(\tau_A) \bm)
	\right)
	\pd{\varphi^{(0)}(x')}{x'^\nu}
	\nonumber\\
	&=
	-\sum_A \pd{}{x^\mu}
	\int d\tau_A^\vph{(\mu}\,
	 S_A^{\lambda(\mu} v_A^{\nu)}
	\bigg[
		\pd{}{x'^\lambda}
	\bigg(
		G\bm( x,x_A(\tau_A) \bm) \pd{\varphi^{(0)}(x')}{x'^\nu}
	\bigg)
	\bigg]_{x'=x_A(\tau_A)},
\end{align}
where $T_A^{\mu\nu}\scount{s}$ refers to the second term in \eqref{eqn:stt_T} and $G(x,x')$ is the retarded Green function for the wave equation, $\Box_x G(x,x')=-\delta^{(4)}(x-x')$. In obtaining the second and third lines, we integrated by parts judiciously and also used the Lorentz invariance of $G(x,x')$ to replace $\partial/\partial x'^\mu \to -\partial/\partial x^\mu$ when it acts on $G(x,x')$. The third line is still rather complicated, but it can be simplified further by taking the nonrelativistic limit. To leading order in~$v$, we have that ${u^\mu_A \sim v_A^\mu}$, and we may replace the integral over~$\tau_A$ with an integral over coordinate time. Additionally, for the retarded Green function, we may use its limiting form, 
$ G(x,x') \sim {\delta(t-t')}/{4\pi|\vec x - \vec x'|}$.
These simplifications leave us with
\begin{equation}
	\varphi^{(1)}\scount{s}(x)
	=
	\sum_A
	\pd{}{x^\mu}
	\bigg[
		\pd{}{x^\lambda}
		\bigg(
			 \frac{ S_A^{\lambda(\mu} v_A^{\nu)} }{4\pi r_A}
			 \pd{\varphi^{(0)}(x')}{x'^\nu}
		\bigg)_{x'=x_A(t)}
		-
		\bigg(
		 	\frac{ S_A^{\lambda(\mu} v_A^{\nu)} }{4\pi r_A}
			\pd{^2\varphi^{(0)}(x')}{x'^\lambda \partial x'^\nu}
		\bigg)_{x'=x_A(t)}
	\bigg].
\end{equation}

Next, we substitute in the solution for $\varphi^{(0)}(x')$, as provided in table~\ref{table:field_profiles}, to leading PN order. Caution must be exercised when doing so, as (after differentiation with respect to~$x'$) evaluating $\varphi^{(0)}$ at the position ${\vec x_A}$ leads to singular self-energy terms. These ultraviolet (power-law) divergences are unphysical, of course, and a more careful analysis shows that they actually vanish in dimensional regularisation, or otherwise can be renormalised away by including appropriate counterterms~\cite{Goldberger:2004jt, Kuntz:2019zef}. In practice, what this means for us is that any singular self-energy term we encounter can simply be discarded. Having done so, we now find
\begin{align}
	\varphi^{(1)}\scount{s}(x)
	=&
	-\sum_A \sum_{B\neq A} \frac{m_A m_B \hat S_A^{ij}}{64\pi^2}
	\bigg[
		\pd{}{x^\mu}\pd{}{x^i}
		\bigg(
			 \frac{ v_A^{\mu} }{r_A}
			 \pd{}{x'^j}\frac{1}{r'_B}
		\bigg)_{x'=x_A(t)}
		\nonumber\\&
		+
		\pd{}{x^i}
		\bigg(
			\frac{ v_A^\mu }{r_A}
			\pd{^2}{x'^\mu \partial x'^j}\frac{1}{r'_B}
		\bigg)_{x'=x_A(t)}
	\bigg],
\label{eqn:disformal_varphi1_deriv}
\end{align}
where we write ${r_B' = |\vec x - \vec x_B(t')|}$. In arriving at this result, we have also made use of \eqref{eqn:gr_SSC}, which tells us that ${S_A^{i0} = S_A^{ij} v_A^j}$; hence, only terms involving $S_A^{ij}$ contribute at leading PN order. Additionally, \eqref{eqn:gr_eom_proper_spin_vectorial} tells us that we can also neglect terms involving time derivatives of ${S^{ij}_A}$ at this order. Evaluating the derivatives in \eqref{eqn:disformal_varphi1_deriv} then leaves us with the final result,
\begin{equation}
	\varphi^{(1)}\scount{s}(x)
	=
	-\sum_A \sum_{B \neq A}
	\frac{3 m_B}{32\pi^2 r_A^2 r^3_{AB} }
	S_A^{ij} n_A^{i \vph{j}} n_{AB}^\avg{jk} v_{AB}^{k \vph{j}}.
\label{eqn:result_varphi_1_S}
\end{equation}
For ease of reference, this formula is also included in table~\ref{table:field_profiles}.

\paragraph{Validity of the ladder expansion.}
These perturbative results are contingent on the ladder expansion in \eqref{eqn:disformal_fields_ladder_expansion} being valid, and so it is worth briefly discussing when this is the case. Roughly speaking, we expect to trust perturbation theory when ${\dis\varphi^{(1)}/\M^4 \ll \varphi^{(0)}}$, and for the spin-independent part of the field profile, this amounts to requiring that the ratio
\begin{equation}
	\frac{\dis \hspace{1pt} \varphi^{(1)}\scount{o}}%
	{\M^4 \varphi^{(0)}\scount{o}}
	\sim
	\frac{\dis \tilde m \tilde v^2}{\M^4 \tilde r^3}
\label{eqn:disformal_breakdown_general}
\end{equation}
is small. This combination is exactly the ladder parameter~$\epsLadder$ defined earlier, apart from the tildes, which we have included in \eqref{eqn:disformal_breakdown_general} to emphasise that our power-counting scheme is in terms of some \emph{characteristic} scales~$\tilde m$, $\tilde v$, and $\tilde r$.

One still has to relate these to the actual physical quantities in the system (like $m_A$, $\vec r_{AB}$, $\vec v_A$, etc.) to determine properly when perturbation theory breaks down, but such a task is by no means straightforward in the general case of $N$~bodies, and so we shall specialise to the more manageable case of a binary system in the remainder of this subsection. When ${N=2}$, we can define
\begin{equation}
	\epsLadder[,1]\scount{o}
	=
	\left|
		\frac{\dis \hspace{1pt} \varphi^{(1)}\scount{o}(\vec x_1)}%
		{\M^4\varphi^{(0)}\scount{o}(\vec x_1)}
	\right|
	\sim
	\frac{\dis m_1 v^2 e^2}{4\pi \M^4 r^3}
\label{eqn:disformal_breakdown_O1}
\end{equation}
as the orbital ladder parameter associated with the first body, with ${r = |\vec x_1 - \vec x_2|}$ denoting the separation of the two bodies, $v$~($\sim\!\sqrt{\GN (m_1 + m_2)/r})$ their average orbital velocity, and $e$ the orbital eccentricity. This last parameter invariably enters into the above expression because $\varphi^{(1)}\scount{o}$ vanishes identically for circular orbits~\cite{Brax:2018bow, Brax:2019tcy, Davis:2019ltc}. The orbital ladder parameter $\epsLadder[,2]\scount{o}$ for the second body then follows by simply interchanging the labels ${1 \leftrightarrow 2}$ in \eqref{eqn:disformal_breakdown_O1}.

If these bodies are spinning, we may also define a pair of spin ladder parameters; namely,
\begin{equation}
	\epsLadder[,1]\scount{s}
	=
	\left|
		\frac{\dis \hspace{1pt} \varphi^{(1)}\scount{s}(\vec x_1)}%
		{\M^4\varphi^{(0)}\scount{o}(\vec x_1)}
	\right|
	\sim
	\left(\frac{m_1}{m_2}\right) \frac{\dis v S_2}{4\pi \M^4 r^4},
\label{eqn:disformal_breakdown_S1}	
\end{equation}
with $\epsLadder[,2]\scount{s}$ similarly given by interchanging the labels ${1 \leftrightarrow 2}$ above. These clear definitions now allow us to assert that our perturbative approach is valid as long as
\begin{equation}
	\max(
		\epsLadder[,1]\scount{o},\, \epsLadder[,2]\scount{o}, \,
		\epsLadder[,1]\scount{s},\, \epsLadder[,2]\scount{s}
	)
	\ll 1.
\label{eqn:disformal_breakdown_condition}	
\end{equation} 

Worth pointing out is the fact that \eqref{eqn:disformal_breakdown_S1} is not suppressed by powers of the eccentricity, since $\varphi^{(1)}\scount{s}$ generically does not vanish for circular orbits. As a result, the spin ladder parameters may well impose stricter conditions on the validity of the ladder expansion for systems with small eccentricities. This is easy to see when the masses and spins of the two bodies are comparable, since in this case we find ${\epsLadder[,1]\scount{o} \sim \epsLadder[,2]\scount{o} \sim e^2 \epsLadder}$ but $\epsLadder[,1]\scount{s} \sim \epsLadder[,2]\scount{s} \,{\sim \epsSpin\hspace{1pt}\epsLadder}$, where $\epsLadder$ is the naive ladder parameter in \eqref{eqn:disformal_breakdown_general}\,---\,having made the identifications ${\tilde m = m_1 = m_2}$, ${\tilde v = v}$, and ${\tilde r = r}$\,---\,from which it follows that ${ \epsLadder[,1]\scount{s} > \epsLadder[,1]\scount{o} }$ when ${\epsSpin > e^2}$.

\subsection{Acceleration}
\label{sec:disformal_acc}

We are now in a position to substitute the field profiles in table~\ref{table:field_profiles} back into~\eqref{eqn:disformal_master}. At 1PM order, the Jordan-frame metric perturbation $\tilde h_{\mu\nu}$ decomposes neatly into a purely general relativistic and purely scalar part, ${ \tilde h_{\mu\nu} = h_{\mu\nu} + \Delta h_{\mu\nu} }$, where the latter reads
\begin{equation}
	\Delta h_{\mu\nu}
	=
	\frac{\con^2}{\Mpl^2}
	\bigg(
		\varphi^{(0)} \eta_{\mu\nu}
		+
		\frac{\dis}{\M^4}
		\Big(
			\varphi^{(1)} \eta_{\mu\nu}
			+
			\partial_\mu\varphi^{(0)}\partial_\nu\varphi^{(0)}
		\Big)
		+
		\mathcal O(\dis^2/\M^8)
	\bigg)
	+
	\mathcal O(\Mpl^{-4}).
\label{eqn:acc_h_Jordan}
\end{equation}
Substituting this into~\eqref{eqn:disformal_master_acc} and discarding any singular self-energy terms in the process [recall the discussion above \eqref{eqn:disformal_varphi1_deriv}] results in an acceleration vector that we can also split into a general relativistic and scalar part, ${\vec a_A^\vph{(} = \vec a_A^\text{(GR)} + \Delta\vec a_A^\vph{(}}$. The former is derived in many standard texts\footnote{See, e.g., eqs.~(9.127), (9.190) and (9.245) of ref.~\cite{PoissonWill} for the end result.} and so need not be revisited here. As for the scalar-induced correction~$\Delta\vec a_A$, its spin-independent part $\Delta\vec a_A\scount{o}$ was also previously discussed in refs.~\cite{Brax:2018bow, Brax:2019tcy, Davis:2019ltc}, and so will not be revisited here either. In what follows, we present our results for the spin-orbit and spin-spin parts.

\paragraph{Spin-orbit terms.}
As per \eqref{eqn:intro_expansion_ladder} and \eqref{eqn:intro_expansion_spin}, we further decompose the spin-orbit acceleration $\Delta\vec a_A\scount{so}$ into its conformal and disformal parts; treating each one in turn. Because ${\varphi^{(0)}\scount{s} = 0}$ at 1PM order for our choice of SSC, the conformal part is simply given by
\begin{align}
	\Delta a_A^{(0)}\scount{so}_i
	&=
	\left(
		\hat S_A^{\mu\nu} v_A^{\lambda \vph{\mu}}
		\partial_\mu \partial_{[i}
		\tilde h_{\lambda]\nu}
	\right)_{\tilde h \to \varphi^{(0)}\scount{o}\eta}
	\nonumber\\
	&=
	\frac{1}{2}\hat S_A^{ij} v_A^{\lambda \vph{j}}
	\partial_\lambda\partial_j \varphi^{(0)}\scount{o}
	\nonumber\\
	&=
	- \sum_{B \neq A}
	\frac{3 m_B }{16\pi r_{AB}^3}
	\hat S_A^{ij}
	n_{AB}^\avg{jk}v_{AB}^{k \vph{j}},
\label{eqn:result_acc_SO_con}
\end{align}
having discarded at each step any term that is subleading in~$v$. Recall also that after differentiating with respect to~$x$, the fields are to be evaluated along the worldline~$\vec x_A(t)$.

As for the disformal part, we find two separate contributions at leading PN order:
\begin{align}
	\Delta a_A^{(1)}\scount{so}_i
	&=
	\frac{1}{2}
	\left(
		\partial_i \tilde h_{00}
	\right)_{\tilde h \to \varphi^{(1)}\scount{s}\eta}
	+\;
	\left(
		\hat S_A^{jk} v_A^{\lambda}
		\partial_j \partial_{[i}
		\tilde h_{\lambda]k}
	\right)_{\tilde h \to \partial\varphi^{(0)}\scount{o}\partial\varphi^{(0)}\scount{o}}
	\nonumber\\
	&=
	-\frac{1}{2} \partial_i\varphi^{(1)}\scount{s}
	-
	\hat S_A^{jk} v_A^{\lambda}
	\big( \partial_i\partial_j\varphi^{(0)}\scount{o} \big)
	\big( \partial_\lambda\partial_k \varphi^{(0)}\scount{o} \big)
	\nonumber\\
	&=
	-
	\sum_{B \neq A} \sum_{C \neq D} \sum_D
	\frac{9 m_B m_C}{ 64\pi^2 r_{AB}^3 r_{CD}^3 }
	n_{AB}^\avg{ij} \hat S_D^{jk}
	n_{DC}^\avg{k\ell} v_{DC}^{\ell \vph{j}}
	(\delta_{AD} - \delta_{BD}).
\label{eqn:result_acc_SO_dis}
\end{align}
It is interesting to note that because the $\varphi^{(1)}\scount{o} \eta_{ij}$ term in the effective metric is suppressed by two powers of~$v$ relative to the $\partial_i\varphi^{(0)}\scount{o}\partial_j\varphi^{(0)}\scount{o} $ term, the former does not contribute to the disformal spin-orbit acceleration at leading PN order. We will find that the same is true also for $\Delta\vec\Omega_A^{(1)}\scount{so}$ later.

\paragraph{Spin-spin terms.}
For the spin-spin terms, the vanishing of $\varphi^{(0)}\scount{s}$ means that ${\Delta\vec a_A^{(0)}\scount{ss} = 0}$ to the order at which we are working, while in the disformal sector, we have~that
\begin{align}
	\Delta a_A^{(1)}\scount{ss}_i
	&=
	\left(
		\hat S_A^{\mu\nu} v_A^{\lambda \vph{\mu}}
		\partial_\mu \partial_{[i}
		\tilde h_{\lambda]\nu}
	\right)_{\tilde h \to \varphi^{(1)}\scount{s}\eta}
	\nonumber\\
	&=
	\frac{1}{2}\hat S_A^{ij} v_A^{\lambda \vph{j}}
	\partial_\lambda\partial_j \varphi^{(1)}\scount{s}
	\nonumber\\
	&=
	\sum_{B\neq A}\sum_{C\neq B}
	\frac{9 m_B m_C }{ 64\pi^2 r_{AB}^3 r_{BC}^3 }
	\hat S_A^{ij}\hat S_B^{k\ell \vph{j}}
	\bigg(
		n_{AB}^\avg{jk} n_{BC}^\avg{\ell p} a_{BC}^{p \vph{\langle}}
		\nonumber\\&\quad
		-
		\frac{5 n_{AB}^\avg{jkq} n_{BC}^\avg{\ell p} v_{AB}^q v_{BC}^p }{r_{AB}}
		-
		\frac{5 n_{AB}^\avg{jk} n_{BC}^\avg{\ell pq}v_{BC}^{pq} }{r_{BC}}
	\bigg).
\label{eqn:result_acc_SS_dis}
\end{align}
Notice here that the 3-acceleration appears also on the rhs of this equation, but since we are working to leading order in $v$, $\epsLadder$, and $\epsSpin$, it suffices to substitute in the Newtonian result,
$\vec a_{BC}^\vph{2} = - \sum_{D\neq B} m_D^\vph{2} \vec n_{BD}^\vph{2}/r_{BD}^2 + \sum_{D\neq C} m_D^\vph{2} \vec n_{CD}^\vph{2}/r_{CD}^2$.

\subsection{Spin precession}
\label{sec:disformal_spin}

We turn now to a derivation of the spin precession rates. As we did with the accelerations, we split the solution to \eqref{eqn:disformal_master_spin} into a general relativistic and scalar part, ${ \vec\Omega_A^\vph{(} = \vec\Omega_A^\text{(GR)} + \Delta\vec\Omega_A^\vph{(} }$; leaving a discussion of the former to standard texts.\footnote{See, e.g., eq.~(9.200) of ref.~\cite{PoissonWill} for the end result.} The scalar-induced correction $\Delta\vec\Omega_A$ is then further decomposed according to \eqref{eqn:intro_expansion_ladder} and \eqref{eqn:intro_expansion_spin}.

\paragraph{Spin-orbit terms.}
In the conformal sector, the spin-orbit precession is set by
\begin{align}
	\Delta\Omega_A^{(0)}\scount{so}^{\hat\imath}
	&=
	\frac{1}{4}\epsilon^{ijk}
	\left(
		v_A^j \partial_k \tilde h_{00}
		-
		2 v_A^\mu\partial_j \tilde h_{k\mu}
	\right)_{\tilde h \to \varphi^{(0)}\scount{o}\eta}
	\nonumber\\
	&=
	\frac{1}{4}\epsilon^{ijk}
	v_A^j\partial_k \varphi^{(0)}\scount{o}
	\nonumber\\
	&=
	\sum_{B\neq A} \frac{m_B}{32\pi r_{AB}^2}
	\epsilon^{ijk} v_A^j n_{AB}^{k \vph{j}},
\label{eqn:result_spin_SO_con}
\end{align}
while in the disformal sector, the contribution at leading PN order comes from 
\bgroup\allowdisplaybreaks
\begin{align}
	\Delta\Omega_A^{(1)}\scount{so}^{\hat\imath}
	&=
	\frac{1}{4}\epsilon^{ijk}
	\left(
		- 2 v_A^\mu\partial_j \tilde h_{k\mu}
	\right)_{\tilde h \to \partial\varphi^{(0)}\scount{o}\partial\varphi^{(0)}\scount{o}}
	\nonumber\\
	&=
	-\frac{1}{2} \epsilon^{ijk}
	\big( v_A^\mu\partial_\mu\partial_j \varphi^{(0)}\scount{o} \big)
	\big( \partial_k \varphi^{(0)}\scount{o} \big)
	\nonumber\\
	&=
	-
	\sum_{B,C\neq A}
	\frac{3 m_B m_C}{128\pi^2 r_{AB}^2 r_{AC}^3 }
	\epsilon^{ijk} n_{AB}^j n_{AC}^\avg{k\ell} v_{AC}^{\ell \vph{\langle}}.
\label{eqn:result_spin_SO_dis}
\end{align}
\egroup

\paragraph{Spin-spin terms.}
The vanishing of $\varphi^{(0)}\scount{s}$ means that ${\Delta\vec\Omega_A^{(0)}\scount{ss} = 0}$ to the order at which we are working, and so the only spin-spin contribution we find is in the disformal sector:
\begin{align}
	\Delta\Omega_A^{(1)}\scount{ss}^{\hat\imath}
	&=
	\frac{1}{4}\epsilon^{ijk}
	\left(
		v_A^j \partial_k \tilde h_{00}
		-
		2 v_A^\mu\partial_j \tilde h_{k\mu}
	\right)_{\tilde h \to \varphi^{(1)}\scount{s}\eta}
	\nonumber\\
	&=
	\frac{1}{4}\epsilon^{ijk}
	v_A^j\partial_k \varphi^{(1)}\scount{s}
	\nonumber\\
	&=
	\sum_{B\neq A}\sum_{C\neq B}
	\frac{9 m_B m_C}{128\pi^2 r_{AB}^3 r_{BC}^3 }
	\epsilon^{ijk} v_A^j \hat S_B^{pq} n_{AB}^\avg{pk}
	n_{BC}^\avg{q\ell} v_{BC}^{\ell \vph{j}}.
\label{eqn:result_spin_SS_dis}
\end{align}

To conclude this section, three final remarks are worth making. First, as we pointed out already in the Introduction, our results for $\Delta\vec a_A^{(0)}$ and $\Delta\vec\Omega_A^{(0)}$ are not new, as they can also be obtained by mapping the conformal part of this scalar-tensor theory onto the PPN formalism~\cite{Will:2018}. Our ability to reproduce these results therefore serves as a good sanity check on our approach. No such map exists between the PPN formalism and the disformal sector, however, and so our results for $\Delta\vec a_A^{(1)}$ and $\Delta\vec\Omega_A^{(1)}$ constitute some of the key contributions of this work.
\looseness=-1

Second, all of our results in this section are valid for any number $N$ of spinning point-like bodies, but we believe it convenient to also have explicit expressions for the special case of a binary system (${N=2}$). In table~\ref{table:summary}, we have thus compiled explicit formulae for the scalar-induced corrections to the binary's relative acceleration, ${\Delta\vec a \equiv \Delta\vec a_1 - \Delta\vec a_2}$, along with the scalar-induced corrections $\Delta\vec\Omega_1$ to the precession rate of the first body's spin vector. (One may then obtain $\Delta\vec\Omega_2$ by simply interchanging the labels ${1 \leftrightarrow 2}$.) In writing these results, the binary's barycentre has been chosen to coincide with the origin such that, in these coordinates, the two worldlines are given by ${\vec x_1 = (m_2/m) \vec r}$ and $\vec x_2 = -(m_1/m) \vec r$, where ${\vec r = \vec x_1 - \vec x_2}$ is their relative separation and ${m=m_1 + m_2}$ is the binary's total mass.

Third, it must be reiterated that we work exclusively with the covariant SSC in~\eqref{eqn:gr_SSC} throughout this paper for convenience. It is nevertheless important to appreciate how these results would be affected had we chosen a different gauge. Because different SSCs correspond to a different choice for the body's effective centre-of-energy coordinate, it can be shown that making the transformation~\cite{Porto:2005ac, Porto:2008tb, Hanson:1974qy, PoissonWill}
\begin{equation}
	\vec x_A \to \vec x_A
	-
	\frac{1}{2m_A}\vec S_A \times \vec v_A + \cdots,
\end{equation}
for instance, takes us from the covariant SSC to the Newton--Wigner one (${m S^{0\mu} + S^{\mu\nu}p_\nu \approx 0 }$). Power counting then reveals that the exact forms of $\Delta\vec a_A^{(0)}\scount{so}$, $\Delta\vec a_A^{(1)}\scount{ss}$, and $\Delta\vec\Omega_A^{(1)}\scount{ss}$ will vary depending on the choice of SSC already at leading PN order, but the remaining equations of motion are invariant at this order.

\section{Gravity Probe~B}
\label{sec:GPB}

In this section, we confront our predictions with spin-precession measurements from the Gravity Probe~B (GPB) experiment~\cite{Everitt:2011hp, Everitt_2015} to establish new constraints on disformally coupled scalar fields.

\paragraph{Drift rates.}
What is measured experimentally are the \emph{drift rates} of a gyroscope\footnote{The GBP satellite actually carries four gyroscopes, but for our purposes it will suffice to focus on just one.} on board the GPB satellite in orbit around the Earth, and so it is useful to begin by defining these quantities. Let $\vec S_\sat$ denote the spin of this gyroscope. Its evolution is governed by \eqref{eqn:intro_eom_spin}, which is difficult to solve exactly, but because the change in its direction ${\vec S_\sat(T)-\vec S_\sat(0)}$ after one orbital period~$T$ is small compared to $\vec S_\sat(0)$, an approximate solution can easily be written down; namely,
\begin{equation}
	\vec S_\sat(T) - \vec S_\sat(0)
	\simeq
	\int_0^T\dx t\; \vec\Omega_\sat(t) \times \vec S_\sat(0).
\end{equation}
Now dividing this equation by~$T$ and the magnitude of $\vec S_\sat$ (which is a conserved quantity\footnote{One can easily see this by taking the dot product of \eqref{eqn:intro_eom_spin} with $\vec S_A$.}), we obtain the drift rate
\begin{equation}
	\dot{\vec s} = \avg{\vec\Omega_\sat \times \vec e_\sat},
\label{eqn:gpb_drift_rate_vector}
\end{equation}
where the unit vector ${\vec e_\sat = \vec S_\sat(0)/S_\sat}$ is aligned along the initial direction of the gyroscope's spin vector, and the angled brackets denote a time average over one orbital period.
Since ${\dot{\vec s} \cdot \vec e_\sat = 0}$, \eqref{eqn:gpb_drift_rate_vector} only has two nontrivial components, which we can project along two orthogonal directions that we shall call North--South~(NS) and East--West~(EW). Mathematically, we~write
\begin{subequations}
\label{eqn:gpb_drift_rate_components}
\begin{align}
	\dot s_\text{NS}
	&\coloneq
	\vec{\dot s} \cdot \vec e_\text{NS}
	=
	\avg{\vec\Omega_\sat \cdot \vec e_\text{EW}},
	\\
	\dot s_\text{EW}
	&\coloneq
	\vec{\dot s}\cdot \vec e_\text{EW}
	=
	- \avg{\vec\Omega_\sat \cdot \vec e_\text{NS}},
\end{align}
\end{subequations}
having chosen ${\vec e_\sat = \vec e_\text{NS} \times \vec e_\text{EW} }$, such that the set $\{ \vec e_\text{NS}, \,\vec e_\text{EW}, \,\vec e_\sat \}$ forms a right-handed basis of orthonormal vectors.

\begin{figure}
\centering\includegraphics{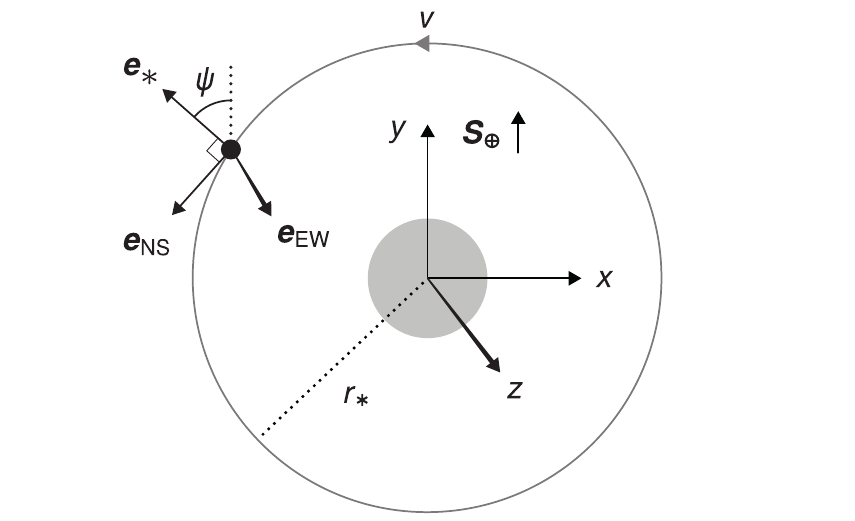}
\caption{Orientation of the Gravity Probe~B satellite relative to the Earth (grey disk). Coordinates are chosen such that the spin~$\vec S_\oplus$ of the Earth points along the $y$~axis, while the orbital angular momentum of the satellite points along the $z$~axis. Centred on the satellite is an orthonormal set of basis vectors, $\{ \vec e_\text{NS}, \, \vec e_\text{EW}, \, \vec e_\sat \}$, with $\vec e_\text{EW}$ chosen to point along the $z$~axis and $\vec e_\sat$ chosen to point in the same direction as the spin $\vec S_\sat$ of the on-board gyroscope at some initial time ${t=0}$.}
\label{fig:GPB}
\end{figure}

The decomposition in \eqref{eqn:gpb_drift_rate_components} is completely general, but the names ``North--South'' and ``East--West'' are particularly appropriate in the context of the GPB experiment, which puts the satellite in a (nearly) circular, polar orbit and orients the gyroscope such that both $\vec S_\sat(0)$ and the Earth's spin vector~$\vec S_\oplus$ lie in the orbital plane; see also figure~\ref{fig:GPB}.
This particular configuration is convenient because, if the vector $\vec e_\text{EW}$ is chosen to genuinely point along the Earth's East--West axis (specifically, in the direction of the satellite's orbital angular momentum vector), then $\vec\Omega_\sat^\text{(GR)}\scount{so}$ contributes only to $\dot s_\text{NS}$, while $\vec\Omega_\sat^\text{(GR)}\scount{ss}$ contributes only to $\dot s_\text{EW}$~\cite{Will:2018}. Such a clean separation enables independent measurements of each effect, which are in good agreement with the predictions of general relativity, as shown in table~\ref{table:GPB_results}.
\bgroup
\setlength{\tabcolsep}{10pt}
\renewcommand{\arraystretch}{1.4}
\begin{table}
\centering
\begin{tabular}{|lcc|}
\hline
& Predicted (mas/yr) & Measured (mas/yr)\\
\hline
$\dot s_\text{NS}$ & $6606.1$ & $6601.8 \pm 18.3$ \\
$\dot s_\text{EW}$ & $39.2$   & $37.2 \pm 7.2$ \\
\hline
\end{tabular}
\caption{Results from the Gravity Probe~B experiment (with their $1\sigma$ uncertainties) and the theoretical predictions from general relativity; reproduced from ref.~\cite{Everitt:2011hp}.  Our conventions for the North--South~(NS) and East--West~(EW) directions follow those of ref.~\cite{Will:2018}.}
\label{table:GPB_results}
\end{table}
\egroup

\paragraph{Constraints on parameter space.}
In our disformal scalar-tensor theory, the scalar field introduces the additional corrections $\Delta \dot s_\text{NS}$ and $\Delta \dot s_\text{EW}$ to the drift rates, whose magnitudes must be bounded from above lest they ruin the good agreement between the general relativistic predictions and the results from GPB.

To compute these scalar-induced corrections, in the rest of this section we shall treat the Earth and the GPB satellite as an isolated two-body system (we comment briefly on the validity of this approximation in appendix~\ref{sec:app_N_bodies}), in which case we find
\begin{subequations}
\begin{align}
	\Delta\vec\Omega_\sat\scount{so}
	&=
	\frac{\con^2}{\Mpl}
	\bigg(
		- \frac{m_\oplus}{32\pi r^3}
		+
		\frac{\dis}{\M^4}
		\frac{m_\oplus^2}{128\pi^2 r^6}
	\bigg)
	\vec\ell,
	\label{eqn:GPB_Omega_SO}
	\\
	\Delta\vec\Omega_\sat\scount{ss}
	&=
	-\frac{\con^2}{\Mpl^2}\frac{\dis}{\M^4}
	\frac{m_\sat}{128\pi^2r^6} \hspace{1pt}
	\vec v \times (\vec v \times \vec S_\oplus),
	\label{eqn:GPB_Omega_SS}
\end{align}
\end{subequations}
where ${\vec\ell = \vec r \times \vec v}$ is the specific orbital angular momentum of the satellite. These results follow from the last two rows of table~\ref{table:summary} after replacing the labels ${(1,2) \to (\sat,\oplus)}$, using the fact that the mass of the satellite~$m_\sat$ is much smaller than the mass of the Earth~$m_\oplus$, and noting that the terms in $\Delta\vec\Omega_\sat\scount{ss}$ proportional to $\vec n \cdot \vec v$ and $\vec\ell \cdot \vec S_\oplus$ vanish due to the circular and polar nature of the orbit.
\looseness=-1

Taking the inner product of these vectors with $\vec e_\text{EW}$ and $\vec e_\text{NS}$, we see that just like in general relativity, the spin--orbit part $\Delta\vec\Omega_\sat\scount{so}$ contributes only to the North--South drift rate~$\Delta\dot s_\text{NS}$, while the spin--spin part $\Delta\vec\Omega_\sat\scount{ss}$ contributes only to the East--West drift rate~$\Delta\dot s_\text{EW}$. Explicitly, we find
\begin{subequations}
\label{eqn:GPB_scalar_drift_rates}
\begin{align}
	\Delta\dot s_\text{NS}
	&=
	\avg{\Delta\vec\Omega_\sat\scount{so}
	\cdot \vec e_\text{EW}}
	=
	- \frac{\con^2 (\GN m_\oplus)^{3/2}}{4 r_\sat^{5/2}}
	\left(
		1 - \frac{\dis m_\oplus}{4\pi \M^4 r_\sat^3}
	\right),
	\label{eqn:GPB_scalar_drift_rates_NS}
	\allowdisplaybreaks\\
	\Delta\dot s_\text{EW}
	&=
	-\avg{\Delta\vec\Omega_\sat\scount{ss}
	\cdot \vec e_\text{NS}}
	=
	\frac{\con^2 \GN^2 m_\oplus}{8 r_\sat^4}
	\left( \frac{\dis m_\sat}{4\pi\M^4 r_\sat^3} \right)
	S_\oplus \sin\psi
	\label{eqn:GPB_scalar_drift_rates_EW}
\end{align}
\end{subequations}
for a circular orbit of radius~$r_\sat$ and velocity ${ |\vec v| = \sqrt{\GN m_\oplus/r_\sat} }$. In obtaining the second line, we defined $\psi$ as the angle between the two spin vectors, $\vec S_\sat(0)$ and $\vec S_\oplus$, and also used two identities:
${
	[\vec v \times (\vec v \times \vec S_\oplus)]^i
	\equiv
	(v^i v^j - \vec v^2 \delta^{ij}) S_\oplus^j
}$
and
${ \avg{v^i v^j} = \vec v^2(\delta^{ij} - \ell^i\ell^j/\vec\ell^2)/2 }$;
the latter following from the fact that $\avg{v^i v^j}$ must be invariant under rotations in the orbital plane.%
\footnote{Treating the satellite's orbit as circular is a good approximation since its eccentricity ${e = 0.0014}$~\cite{GPBreport}, but one can nevertheless show that \eqref{eqn:GPB_scalar_drift_rates_NS} generalises to
\begin{equation*}
	\Delta\dot s_\text{NS}
	=
	- \frac{ \con^2 (\GN m_\oplus)^{3/2} }{4 r_\sat^{5/2}(1-e^2)}
	\bigg(
		1 -
		\frac{\dis m_\oplus}{4\pi\M^4 r_\sat^3} 
		\frac{1 + 3e^2 + 3e^4/8}{(1-e^2)^3}
	\bigg)
\end{equation*}
for general values of ${e \in [0,1)}$, while still assuming a polar orbit (${\vec\ell\cdot \vec S_\oplus = 0}$). This result follows from direct evaluation after writing $|\vec\ell|^2 = \GN m_\oplus r_\sat (1-e^2)$ and ${r = r_\sat(1 - e \cos\mathcal E)}$, where the eccentric anomaly $\mathcal E$ evolves in time according to the differential equation $\dx t/\dx\mathcal E = T(1 - e \cos\mathcal E)/(2\pi)$~\cite{PoissonWill}. Although more challenging, one could in principle use the same procedure to generalise $\Delta\dot s_\text{EW}$ to the case of eccentric orbits as well.}

\bgroup
\setlength{\tabcolsep}{10pt}
\renewcommand{\arraystretch}{1.2}
\begin{table}
\centering
\begin{tabular}{|lcl|}
\hline
Mass of the Earth & $m_\oplus$ & $5.9724 \times 10^{24}$~kg
\\
Mean radius of the Earth & $R_\oplus$ & 6371~km
\\
Spin angular momentum  of the Earth & $S_\oplus$ & 
$2.412 \times 10^{-5}~m_\oplus^\vph{2}\,R_\oplus^2~\text{s}^{-1}$
\\[1pt]
\hline
Mass of the satellite & $m_\sat$ & 3100~kg
\\
Semi-major axis of the satellite's orbit & $r_\sat$ & 7027.4~km
\\
Orbital eccentricity & $e$ & 0.0014
\\
Angle between $\vec S_\sat(0)$ and $\vec S_\oplus$ & $\psi$ & $73.2^\circ$
\\[1pt]
\hline
\end{tabular}
\caption{Parameters relevant to the Gravity Probe~B experiment; reproduced from refs.~\cite{Will:2018, GPBreport}.}
\label{table:GBP_parameters}
\end{table}
\egroup

Now confronting \eqref{eqn:GPB_scalar_drift_rates_NS} with the data in table~\ref{table:GPB_results} gives us the $2\sigma$ constraint
\begin{equation}
	| (6606.1 + \Delta \dot s_\text{NS}) - 6601.8 | < 36.6,
\label{eqn:GPB_constraint_NS}
\end{equation}
which (after using the numerical values in table~\ref{table:GBP_parameters}) may be written more transparently as%
\footnote{Note that the $\mathcal{O}(\dis)$ term in \eqref{eqn:GBP_bound_NS} stems from the $\partial_\mu \varphi^{(0)} \partial_\nu \varphi^{(0)}$ term in $\Delta h_{\mu\nu}$ [cf.~\eqref{eqn:acc_h_Jordan} and~\eqref{eqn:result_spin_SO_dis}]. This can be much larger than the $\mathcal{O}(\dis^{\,0})$ term without violating the validity of the ladder expansion, as the $\mathcal{O}(\dis^{\,2})$ corrections to \eqref{eqn:GBP_bound_NS} are still guaranteed to be suppressed when ${\dis\varphi^{(1)}/\M^4 \ll \varphi^{(0)}}$.}
\begin{equation}
	\left|\,
		4.3
		-
		1.1 \times 10^3 \,\con^2
		+
		8.9 \times 10^{17}\,
		\con^2\dis
		\left( \frac{9.2~\text{eV}}{\M}\right)^4
	\right|
	< 36.6.
\label{eqn:GBP_bound_NS}
\end{equation}
The middle term on the lhs is negligible for values of $\con^2$ ($< 2.5 \times 10^{-5}$) not already ruled out by the Cassini spacecraft~\cite{Bertotti:2003rm}, and so in this region of parameter space, \eqref{eqn:GBP_bound_NS} reduces to the upper bound on $|\con^2\dis/\M^4|$ as given in~\eqref{eqn:intro_bound_GPB}. (At the order-of-magnitude level, the upper bound is the same for both positive and negative values of~$\dis$.) Of course, caution must be exercised when interpreting this upper bound, as our perturbative results are valid only when \eqref{eqn:disformal_breakdown_condition} holds. For this system, the largest ladder parameter turns out to be $\epsLadder[,\oplus]\scount{o}$, and requiring that this remain small translates into the condition ${|\dis|^{-1/4}\M \gg 9.2~\text{eV}}$, as per~\eqref{eqn:intro_bound_GPB_ladder_cutoff}.%
\footnote{%
To check that $\epsLadder[,\oplus]\scount{s}$ and $\epsLadder[,\sat]\scount{s}$ are indeed smaller than $\epsLadder[,\oplus]\scount{o}$, we used the information in ref.~\cite{Bencze:2015uij} to arrive at the estimate $S_\sat \approx 3.84~\text{g}\,\text{m}^2\,\text{s}^{-1}$ for the spin of the on-board gyroscope.
As for the regimes of validity in \eqref{eqn:intro_bound_Mercury} and \eqref{eqn:intro_bound_Saturn}, those in  \eqref{eqn:intro_bound_Mercury_expansion} and \eqref{eqn:intro_bound_Saturn_expansion} follow from requiring that the Sun has orbital ladder parameter ${\epsLadder[,\odot]\scount{o} \ll 1}$, while the regimes in \eqref{eqn:intro_bound_Mercury_resum} and \eqref{eqn:intro_bound_Saturn_resum}, wherein the ladder expansion can be resummed, follow from requiring that $\epsLadder[,\odot]\scount{o}\epsLadder[,\text{P}]\scount{o}\gg 1$~\cite{Davis:2019ltc}, with ``P'' a placeholder for the appropriate planet.}
Taking both of these inequalities into account, the region of parameter space that we can confidently exclude is shaded in red in figure~\ref{fig:constraints_M}.%

Notice that we have yet to include the effect of \eqref{eqn:GPB_scalar_drift_rates_EW}, but it is easy to see that this provides no added improvement to our constraints, as $\Delta\dot s_\text{EW}$ is suppressed relative to the disformal term in~$\Delta\dot s_\text{NS}$ by the small  mass ratio~$m_\sat/m_\oplus$, two powers of the orbital velocity~$v$, and one power of ${S_\oplus/|m_\oplus\vec\ell|}$. Indeed, plugging in the numbers shows that ${\Delta\dot s_\text{EW} = 2.6 \times 10^{-15}\, \con^2\dis\, (9.2~\text{eV}/\M)^4}$. This is just as well, since (as we discussed at the end of section~\ref{sec:disformal}) the precise value of the scalar-induced spin-spin precession rate $\Delta\vec\Omega_\sat\scount{ss}$ depends on the choice of SSC already at leading PN order, while the scalar-induced spin-orbit precession rate $\Delta\vec\Omega_\sat\scount{so}$ does not. The fact that the GPB constraint in figure~\ref{fig:constraints_M} utilises only the latter therefore means that this result is independent of our choice of SSC.

\paragraph{Constraints on the Vainshtein radii.}
Although bounding the size of $\Delta\vec\Omega_\sat\scount{ss}$ does not lead to any meaningful constraint, it is worth briefly remarking that a would-be measurement of this effect provides qualitatively different information than a measurement of $\Delta\vec\Omega_\sat\scount{so}$. The former may be regarded as probing the effective Vainshtein radius of the satellite (but note that this statement must be refined once we take the other masses present in the Solar System into account; see appendix~\ref{sec:app_N_bodies} for more details), while the latter probes the effective Vainshtein radius of the Earth.

\begin{figure}
\centering
\includegraphics{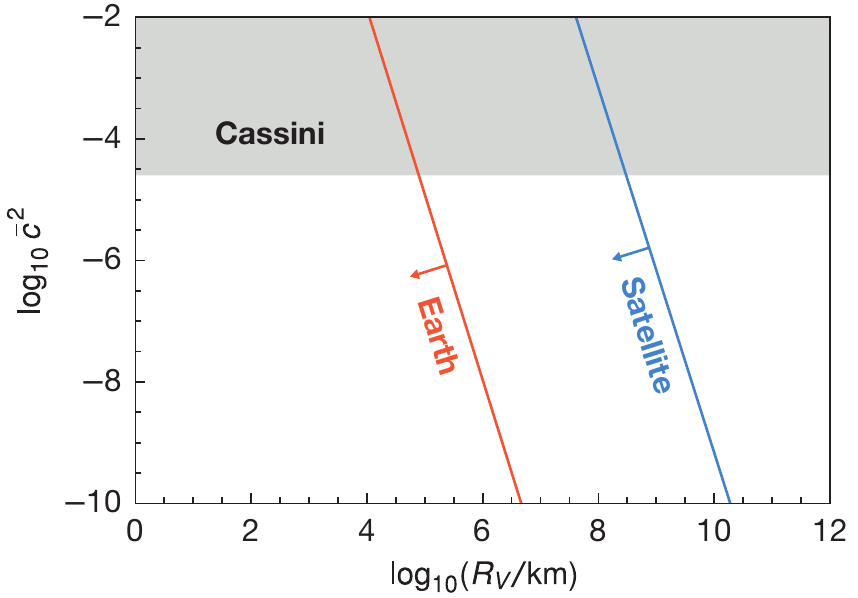}
\caption{The Gravity Probe~B (GPB) experiment constrains the Vainshtein radii~$R_V$ of the Earth and of the GPB satellite to be in the region to the left of the red and blue solid lines, respectively (having set ${\dis=1}$). Shaded in grey is the region of parameter space excluded by the Cassini spacecraft~\cite{Bertotti:2003rm}.}
\label{fig:constraints_RV}
\end{figure}

To see this, we simply rewrite the disformal parts of \eqref{eqn:GPB_Omega_SO} and \eqref{eqn:GPB_Omega_SS} as
\begin{subequations}
\begin{align}
	\frac{\dis}{\M^4}\Delta\vec\Omega_\sat^{(1)}\scount{so}
	&=
	\frac{\dis R_V^3(m_\oplus)}{r^3}
	\frac{m_\oplus \vec\ell}{32\pi r^3},
	\\
	\frac{\dis}{\M^4}\Delta\vec\Omega_\sat^{(1)}\scount{ss}
	&=
	-\frac{\dis R_V^3(m_\sat)}{r^3}
	\frac{\vec v \times (\vec v \times \vec S_\oplus)}{32 \pi r^3} \hspace{1pt},
\end{align}
\end{subequations}
where $R_V(m) = [m/(4\pi \M^4)]^{1/3}$ is the Vainshtein radius associated with the mass~$m$~\cite{Davis:2019ltc}. Confronting these expressions with the data in~table~\ref{table:GPB_results} then gives us the upper bounds on $R_V(m_\oplus)$ and $R_V(m_\sat)$ as shown in figure~\ref{fig:constraints_RV}. By definition, we expect ${R_V(m_\sat) \ll R_V(m_\oplus)}$, and so the fact that  GPB places a tighter constraint on $R_V(m_\oplus)$ than it does on $R_V(m_\sat)$ is just another way of illustrating that bounding the size of $\Delta\dot s_\text{EW}$ establishes no meaningful constraint on the parameters of our model.

\section{Discussion}
\label{sec:disc}

In this work, we have shown how disformally coupled scalar fields influence the evolution of spinning point-like bodies. Working directly at the level of the equations of motion, we first demonstrated that because rotational energy necessarily gravitates, the inclusion of spin generically introduces new terms into the scalar-field profile sourced by these objects, but at first post-Minkowskian order, such terms appear only in the disformal sector when imposing the covariant spin supplementary condition. Staying in this particularly convenient gauge, we then quantified the scalar's impact on the accelerations of these bodies and on the rates at which their spins precess\,---\,deriving compact expressions at leading order in a nonrelativistic, post-Newtonian expansion; first for the general case of a system with $N$ spinning point-like bodies, and then for the special case of a binary system with ${N=2}$. Confronting these results with spin-precession measurements from the Gravity Probe~B experiment then allowed us to place new (gauge-invariant) constraints on disformally coupled scalar fields, which we found to be much stronger than existing Solar~System bounds from perihelion precession. This work therefore establishes spin effects as a new and promising avenue along which to search for nonminimally coupled, light scalar fields. More work is needed to fully realise this potential, and so we conclude by identifying several key directions in which to extend our results.

\paragraph{Going beyond the ladder expansion.}
Because the validity of our perturbative solutions rely on the ladder parameter(s) being small, we were able to use the Gravity Probe~B experiment to place constraints only in the region ${|\dis|^{-1/4}\M \gg 9.2~\text{eV}}$. To probe even smaller values of this disformal coupling scale would require resumming our results to all orders in~$\dis/\M^4$. As we discussed already in the Introduction, such a procedure was recently shown to be possible for  binary systems of nonspinning bodies~\cite{Davis:2019ltc}, and it seems likely that a similar approach could be applied also to the spinning case.

\paragraph{Constraints from the orbital motion.}
The constraints we established in figure~\ref{fig:constraints_M} utilise only our predictions for the scalar-induced corrections $\Delta\vec\Omega_A$ to the spin-precession rate, but in section~\ref{sec:disformal} we showed that a disformal interaction also induces novel spin--orbit and spin--spin effects in the orbital motion of a binary. Within the Solar System, laser-ranging data of the LAGEOS and LARES satellites have been used to test the general relativistic prediction for spin--orbit acceleration at the 5\% level~\cite{Ciufolini:2016ntr}, but these results do not place any meaningful constraints on the disformal coupling, since the scalar-induced correction $\Delta\vec a^{(1)}\scount{so}$ is proportional to ${\vec S_+ = (m_2/m_1)\vec S_1 + (m_1/m_2)\vec S_2}$, which vanishes for satellites of negligible mass and spin relative to the Earth. It would therefore be interesting to explore if ranging measurements of a future satellite with nonnegligible spin could be used to place meaningful constraints on the disformal coupling. Satellite missions aside, it would also be interesting to explore how well disformal spin-orbit effects can be constrained by pulsar binaries, since these systems have constituents with comparable masses and spins.

\paragraph{\emph{N}-body effects.}
To translate the measurements from Gravity Probe~B into a concrete bound on the disformal coupling, we assumed for simplicity that the Earth--satellite system could be treated as an isolated binary. In appendix~\ref{sec:app_N_bodies}, we used rough estimates to argue that this was indeed a valid approximation for the scalar-induced spin-orbit precession rate~$\Delta\vec\Omega_A\scount{so}$, but found that the spin-spin precession rate~$\Delta\vec\Omega_A\scount{ss}$ was highly sensitive to the presence of the other masses in the Solar System (primarily the Sun). While this ultimately did not affect the end result in figure~\ref{fig:constraints_M} due to $\Delta\vec\Omega_A\scount{ss}$ still being orders of magnitude smaller than $\Delta\vec\Omega_A\scount{so}$, it would be interesting to explore if other astrophysical systems could exhibit a strong sensitivity to disformal $N$-body effects.

\paragraph{Calculations in the Jordan frame.}
We have here chosen to carry out our analysis in the Einstein frame, wherein metric fluctuations have a canonical kinetic term (the Einstein--Hilbert term of general relativity), but where matter couples nonminimally to an effective metric containing a light scalar field. Upon a suitable field redefinition, this theory can be recast into the Jordan frame, in which matter couples minimally to a metric at the cost of nontrivial kinetic terms in the scalar-tensor sector. Given that this kind of field redefinition leaves physical observables unaffected, it should be possible to re-derive our results in the Jordan frame. Such an exercise could shed light on how spinning bodies evolve in modified theories of gravity, like in Horndeski scalar-tensor theories~\cite{Horndeski:1974wa}.

\paragraph{Connecting to cosmology.}
The simple toy model of a massless scalar used in this paper serves to illustrate the kinds of effects that disformal couplings can have on spinning point-like bodies, but a more sophisticated scalar-tensor sector must be included if it is to be a phenomenologically viable candidate for dark matter or dark energy. It would therefore be interesting to study the interplay between the Solar System-scale constraints presented here and the cosmological constraints that have been established on much larger scales~\cite{Koyama:2015vza, Joyce:2016vqv, Creminelli:2017sry, Ezquiaga:2017ekz, Sakstein:2017xjx, Baker:2017hug, Heisenberg:2018vsk, Ezquiaga:2018btd, Kase:2018aps, deRham:2018red, Baker:2019gxo}.

\paragraph{Gravitational and scalar waves.}
By including the effects of spin, our results offer a broader picture of how light scalar fields influence the evolution of point-like bodies, but this picture is not yet complete as our analysis was here restricted to a purely conservative setting. Equally important now in this era of gravitational-wave astronomy is an understanding of radiative effects; hence, it will be interesting in the future to study how spin enters into a binary's radiative multipoles in disformal scalar-tensor theories.

\acknowledgments
It is a pleasure to thank Joel Berg\'e and Jeremy Sakstein for helpful discussions.
This work was partially supported by STFC Consolidated Grants No.~ST/P000673/1, No.~ST/P000681/1, and No.~ST/T000694/1. S.M. is supported by an Emmanuel College Research Fellowship.
This work was made possible by the Institut Pascal at Universit\'e Paris--Saclay with the support of the P2I and SPU research departments and the P2IO Laboratory of Excellence (program ``Investissements d'avenir'' ANR-11-IDEX-0003-01 Paris-Saclay and ANR-10-LABX-0038), as well as by the IPhT.

\appendix
\section{Spin precession with \emph{N} bodies}
\label{sec:app_N_bodies}

Our results in section~\ref{sec:GPB} assume for simplicity that the Earth and the Gravity Probe~B (GPB) satellite can be treated as an isolated two-body system. In this appendix, we assess the validity of this assumption by estimating the size of the corrections to $\Delta\vec\Omega_\sat\scount{so}$ and $\Delta\vec\Omega_\sat\scount{ss}$ coming from the other masses in the Solar System. To obtain rough order-of-magnitude estimates for these $N$-body effects, we will ignore the vectorial nature of these quantities and will power count as follows:
\begin{equation}
	\epsilon^{ijk} \sim 1,
	\quad
	n_{AB}^i \sim 1,
	\quad
	v_{AB}^i \sim v_{AB} \,(\equiv\!| \vec v_{AB}|),
	\quad
	S_A^{i} \sim S_{A}^{ij} \sim S_A \,(\equiv\!|\vec S_A|).
\end{equation}

Applying these rules to \eqref{eqn:result_spin_SO_con} and \eqref{eqn:result_spin_SO_dis}, we find that the conformal and disformal parts of $\Delta\vec\Omega_\sat\scount{so}$ scale like
\begin{align}
	\Delta\Omega_\sat^{(0)}\scount{so}
	&\sim
	\frac{m_\oplus v_{\sat\oplus}}{r_\sat^2}
	\bigg\{
		1
		+ \sum_{ A \notin \{\sat,\oplus\} }\frac{F_{\sat A}}{F_{\sat\oplus}}
	\bigg\},
	\label{eqn:app_SO_con}
	\\
	\Delta\Omega_\sat^{(1)}\scount{so}
	&\sim
	\frac{m_\oplus^2 v_{\sat\oplus}}{r_\sat^5}
	\bigg\{
		1
		+
		\sum_{ A \notin\{\sat,\oplus\} } \frac{F_{\sat A}}{F_{\sat\oplus}}
		\bigg( 1 + \frac{\omega_{\sat A}}{\omega_{\sat\oplus}} \bigg)
		+
		\sum_{A,B \notin \{\sat,\oplus\}}
		\frac{F_{\sat A}}{F_{\sat\oplus}} \frac{F_{\sat B}}{F_{\sat\oplus}}
		\frac{\omega_{\sat B}}{\omega_{\sat\oplus}}
	\bigg\},
	\label{eqn:app_SO_dis}
\end{align}
where the sums in the curly brackets run over all of the other bodies whose effects we wish to take into account. Notice that these sums are expressed in terms of just two types of quantities: the (magnitude of the) Newtonian force ${ F_{AB} \sim \GN m_A m_B / r_{AB}^2 }$ between the objects $A$ and~$B$, and the effective angular velocity ${ \omega_{AB} \sim v_{AB}/r_{AB} }$ for the motion of $A$ relative to $B$.

The question of whether the Earth--satellite system is, to a good approximation, an isolated two-body system can now be translated into a question of whether these sums are small. For a concrete example, let us estimate the effect of the Sun on $\Delta\vec\Omega_\sat\scount{so}$. Taking the distance between the Sun and the GPB satellite, $r_{\sat\odot}$, to be approximately equal to the distance between the Sun and the Earth, we find that the ratio of forces evaluates~to
\begin{equation}
	\frac{F_{\sat A}}{F_{\sat\oplus}}\bigg|_{A=\odot}
	\sim 
	\frac{m_\odot}{m_\oplus}
	\left( \frac{ r_\sat }{ r_{\sat\odot} }\right)^2
	\sim
	7 \times 10^{-4}.
\end{equation}

Similarly, we find that the ratio of angular velocities is
\begin{equation}
	\frac{\omega_{\sat A}}{\omega_{\sat\oplus}}\bigg|_{A=\odot}
	\sim
	\frac{v_{\sat\odot}}{v_{\sat\oplus}} \frac{r_{\sat}}{r_{\sat\odot}}
	\sim
	2 \times 10^{-4}
\end{equation}
after assuming that ${v_{\sat\odot} \sim v_{\sat\oplus} + v_{\oplus\odot}}$, where $v_{\sat\oplus}$ is the orbital velocity of the satellite around the Earth while $v_{\oplus\odot}$ is the orbital velocity of the Earth around the Sun. Of course, the true value of $v_{\sat\odot}$ will be smaller than this estimate since the vectors $\vec v_{\sat\oplus}$ and $\vec v_{\oplus\odot}$ need not be parallel to one another, but taking this naive sum gives us an upper bound on the size of $v_{\sat\odot}$.

Analogous estimates can be made for the other bodies in the Solar System (e.g., the moon, the other planets, etc.), and in doing so we find that the Sun provides the largest such contribution to the sums in \eqref{eqn:app_SO_con} and \eqref{eqn:app_SO_dis}; hence, the Earth--satellite system can indeed, to a good approximation, be treated as an isolated binary when calculating $\Delta\vec\Omega_\sat\scount{so}$.

As we will now argue, the same is not true for $\Delta\vec\Omega_\sat\scount{ss}$. The same power counting rules as before tell us that the disformal spin-spin precession rate in \eqref{eqn:result_spin_SS_dis} scales like
\begin{align}
	\Delta\Omega_\sat^{(1)}\scount{ss}
	\sim
	\frac{m_\sat v_{\sat\oplus}^2 S_\oplus}{ r_\sat^6 }
	\bigg\{
		1
		+
		\sum_{A\notin\{\sat,\oplus\}}
		\frac{F_{A\oplus}}{F_{\sat\oplus}}
		\frac{\omega_{A\oplus}}{\omega_{\sat\oplus}}
		+
		\sum_{A\notin\{\sat,\oplus\}} \sum_{B\neq A}
		\frac{S_B}{S_\oplus}
		\frac{m_B}{m_\sat}\frac{\omega_{AB}}{\omega_{\sat\oplus}}
		\frac{r_\sat^5}{r_{\sat A}^3 r_{AB}^2}
	\bigg\},
\end{align}
and we find that the largest term in the curly brackets, coming from the Sun, is
\begin{equation}
	\frac{F_{A\oplus}}{F_{\sat\oplus}}
	\frac{\omega_{A\oplus}}{\omega_{\sat\oplus}}
	\bigg|_{A=\odot}
	\sim
	\frac{m_\odot}{m_\sat}
	\bigg( \frac{r_\sat}{r_{\odot\oplus}} \bigg)^3
	\frac{v_{\odot\oplus}}{v_{\sat\oplus}}
	\sim
	3 \times 10^{14}.
\label{eqn:app_SS_sun_term}
\end{equation}

What does all of this mean for our results in the main text? First, as we pointed out already, the fact that including the other masses in the Solar System leads to only fractional corrections in $\Delta\vec\Omega_\sat\scount{so}$ guarantees that the constraint we established from bounding the size of $\Delta\dot s_\text{NS}$ is trustworthy. Second, while including the orbit of the Sun does increase the size of $\Delta\vec\Omega_\sat\scount{ss}$ by many orders of magnitude, even then, the ratio
\begin{equation}
	\frac{\Delta\Omega_\sat^{(1)}\scount{ss}}%
	{\Delta\Omega_\sat^{(1)}\scount{so}}
	\sim
	\frac{m_\sat v_{\sat\oplus}^2 S_\oplus / r_\sat^6}%
	{ m_\oplus^2 v_{\sat\oplus} / r_\sat^5 }
	\bigg\{
		\frac{F_{\odot\oplus}}{F_{\sat\oplus}}
		\frac{\omega_{\odot\oplus}}{\omega_{\sat\oplus}}
	\bigg\}
	\sim
	2 \times 10^{-18}
\end{equation}
is still a very small number, meaning $\Delta\dot s_\text{EW}$ still does not establish any meaningful constraint on the parameter space of our theory. Taken together, these two points tell us that our results in figure~\ref{fig:constraints_M} are fully robust against corrections from $N$-body effects.

The one conclusion that does change in the main text is our interpretation of $\Delta\vec\Omega_\sat^{(1)}\scount{ss}$ as being a probe of the satellite's Vainshtein radius. Assuming that our crude estimate in \eqref{eqn:app_SS_sun_term} holds (i.e., that this effect from the Sun is not found to vanish due to time-averaging or a coincident alignment of vectors when undertaking a detailed calculation) and so dominates over the term in \eqref{eqn:GPB_Omega_SS}, a measurement of $\Delta\vec\Omega_\sat^{(1)}\scount{ss}$ would thus provide a means of probing the Vainshtein radius of the Sun.

\bibliographystyle{JHEP-custom}
\bibliography{disformal_spin}
\end{document}